\documentclass[12pt,draftcls,  onecolumn, twoside]{IEEEtran}

\usepackage[utf8]{inputenc} 
\usepackage[T1]{fontenc}
\usepackage{url}
\usepackage{ifthen}
\usepackage{cite}
\usepackage[cmex10]{amsmath} 
\usepackage{amssymb}
\usepackage{graphicx}
\usepackage{wrapfig}
\usepackage{multirow}
\usepackage{array}
\usepackage{makecell}
\usepackage{tikz}
\usepackage{algorithm}
\usepackage{algpseudocode}
\usepackage{comment}
\usepackage[english]{babel}
\usepackage{fancyhdr}
\usepackage{epstopdf}
\newtheorem{thm}{Theorem}

\newtheorem{prop}{Proposition}

\newtheorem{exmp}{Example}

\newtheorem{rem}{Remark}

\begin{document}
	\title{Security and Privacy in Cache-Aided Linear Function Retrieval for Multi-access Coded Caching    } 
	
	\author{%
		\IEEEauthorblockN{Mallikharjuna Chinnapadamala and B. Sundar Rajan }\\
		\IEEEauthorblockA{Department of Electrical Communication Engineering, Indian Institute of Science, Bengaluru 560012, KA, India \\
			E-mail: \{chinnapadama,bsrajan\}@iisc.ac.in}
	\thanks{Part of the content of this manuscript appears in proceedings of \textit{ IEEE Information Theory Workshop(ITW) November 01-09, 2022}, Mumbai \cite{ChR}.}}
	\maketitle
	\begin{abstract}
		A multi-access network  consisting of $N$ files, $C$ caches, $K$ users with each user having access to a unique set of $r$  caches has been introduced recently by Muralidhar et al. ("Maddah-Ali-Niesen Scheme for Multi-access Coded Caching," in \textit{Proc. ITW}, 2021). It considers Single File Retrieval (SFR)  i.e, each user demands an arbitrary file from the server. It proposes a coded caching scheme  which was shown to be optimal under the assumption of uncoded placement by Brunero and Elia ("Fundamental Limits of Combinatorial Multi-Access Caching" in {\textit{arXiv:2110.07426} }). The above multi-access network is referred to as combinatorial topology which is considered in this work with three additional features : a) Linear Function Retrieval (LFR) i.e., each user is interested in retrieving an arbitrary linear combination of files in the server's library; b) Security i.e., the content of the library must be kept secure from an eavesdropper who obtains the signal sent by the server; c) Privacy i.e., each user can only get its required file and can not get any information about the demands of other users. Achievable Secure, Private LFR (SP-LFR) scheme, Secure LFR (S-LFR) scheme and Improved S-LFR scheme are proposed. As special cases, our work recovers some of the results by Yan and Tuninetti ("Key Superposition Simultaneously Achieves Security and Privacy in Cache-Aided Linear Function Retrieval," in \textit{Trans. Inf. Forensics and Security}, 2021")  and Sengupta et al.("Fundamental limits of caching with secure delivery," in \textit{Trans. Inf. Forensics and Security}, 2015). At a memory point, $M=\frac{r\binom{C}{r}}{C}$, the SP-LFR scheme is within a constant multiplicative factor from the optimal rate for $N\geq2Kr$ and at, $M=\frac{\binom{C}{r}}{C}$, the improved S-LFR scheme is within a constant multiplicative factor from the optimal rate for $N\geq2K$.

	\end{abstract}
	\section{INTRODUCTION}
	\label{sec1}	
	Coded caching is a promising technique to reduce network congestion during peak traffic hours by duplicating content at the users. Coded caching was first proposed by Maddah-Ali and Niesen (MAN) \cite{MaN}. It operates in two phases: (1) a placement phase, where each user's cache is populated to its size, and (2) a delivery phase, where users reveal their demands and the server has to satisfy their demands. In this phase, the server exploits the content of the caches to reduce network traffic. It was shown in \cite{MaN}  that it is possible to achieve "global caching gain" by serving several users simultaneously with single transmission. But, subpacketization (the number of sub files each file is split into) levels increase exponentially with the total number of users in the MAN scheme. The construction of coded caching schemes having lower subpacketization levels  based on the Placement Delivery Arrays (PDA) was given in \cite{YCTC}. Coded caching schemes from linear block codes \cite{TaR} and block designs \cite{KrP1} were reported and these have practically achievable subpacketization levels. 
            	
	A network is called a dedicated cache network when each user has access to a dedicated cache. Multi-access networks are those networks where each user can access multiple caches. These type of networks are considered in \cite{RaK},\cite{SPE},\cite{MKR},\cite{HKD},\cite{KMR}. A coded caching scheme for multi-access networks with a number of users much larger than the number of caches and access ratio $r$, i.e., each user is having access to a unique set of $r$ caches was proposed in \cite{MKR}. A multi-access network with $K$ users and $K$ caches, with each user having access to $r$ caches in a cyclic wrap-around way is considered in \cite{HKD}. Some of the works \cite{HKD},\cite{NiM},\cite{DeD} considered networks where demands are non-uniform i.e., all files are not equally likely to be requested by each user. However, we are restricting ourselves to the case where all files are equally likely to be requested by each user. The problem of linear function retrieval is considered in \cite{WSJTC} and it studies the scenario where users are interested in retrieving a linear function of files and not just a single file. It is shown that the rate (number of files sent by the server to satisfy the users demands) for cache-aided linear function retrieval depends on the number of linearly independent functions that are demanded. When all the demanded functions are linearly independent, then the rate achieved for linear function retrieval is same as the rate of MAN scheme for single file retrieval. A class of linear functions that operate element-wise over files is considered in \cite{WSJTC}, whereas \cite{WSJTC2} considers a more general case where users request general linear combinations of all symbols in the library. However, we restrict ourselves to the case considered in \cite{WSJTC}. We refer to the schemes where each user is allowed to demand an arbitrary file from the library as Single File Retrieval (SFR) schemes and schemes where each user demands an arbitrary linear combination of the files as Linear Function Retrieval (LFR) schemes. Demand privacy for SFR was studied in \cite{WaC}, \cite{Kam}, \cite{AST}, where a user should not gain any information about the index of the file demanded by another user. Demand privacy against colluding users along with LFR was considered in \cite{YaT2}. Content security for SFR was studied in \cite{SeT}, where the system needs to protect the content of the library against an eavesdropper who obtains the signal sent by the server in the delivery phase and we refer to it as a security key scheme. In addition, the linear function retrieval with privacy against colluding users and security from external eavesdropper is considered in \cite{YaT}. The idea of key superposition to guarantee both content security and demand privacy simultaneously was used in \cite{YaT}. Demand privacy  using privacy keys for the multi-access networks where each user has access to $r$ caches in a cyclic wrap-around way \cite{HKD} is considered in \cite{NaR},\cite{LWCC}. But, the multi-access network we consider has number of users much larger than the number of caches. It consists of a server having $N$ files which is connected to $K$ users, where each user has access to a unique set of $r$ caches out of $C$ caches. Demand privacy for single file retrieval is considered in \cite{NaR},\cite{LWCC} whereas we consider security and demand privacy for linear function retrieval.
	
	In the following subsections, we give a brief review of Maddah-Ali-Niesen(MAN) scheme for the multi-access coded caching and Shamir's secret sharing technique.
	\subsection{Maddah-Ali-Niesen Scheme for Multi-access Coded Caching \cite{MKR}}
	 A multi-access network consisting of $C$ caches and $K$ users, with each user having access to a unique set of $r$ caches was considered in \cite{MKR}. This network can accommodate a large number of users at low subpacketization levels. This network is referred to as combinatorial topology, and the scheme given in \cite{MKR} was proved to be optimal under the assumption of uncoded placement in \cite{BrE}. The well-known Maddah-Ali-Niesen(MAN) scheme becomes a special case of the scheme proposed in \cite{MKR} when $r=1$. The problem setup is as follows: The server containing $N$ files denoted as $ W_{1}, W_{2}, W_{3}....W_{N}$  each of size $F$ bits is connected to $k$ users through an error-free shared link and each user is having access to a unique set of $r$ caches out of $C$ caches. Each cache is of size $M$ files, and $Z_{c}$ denotes the content of cache $c\in[C]$. During the placement phase, the server divides each file into $\binom{C}{t}$ sub files where $t=\frac{CM}{N}$ and $c^{th}$ cache is filled as follows 
	\begin{center}
	$Z_{c} = \{W_{i,\mathcal{T}} : c\in\mathcal{T}, \mathcal{T} \subset C, |\mathcal{T}| =t \quad \forall \quad i \in [N] \}$
	\end{center}
	Let $d_{U}$, where $U \subset C$, $|U|=r$ denote the demand of the user connected to the set of caches represented by $U$. Then, for each $S \subset C$, $|S|=t+r$ server transmits
		$\underset{\substack{U\subset S\\|U|=r}}{\bigoplus} W_{{{d_{U}}},S\backslash U}$.
	 Thus, the scheme achieves a rate, $R=\frac{\binom{C}{t+r}}{\binom{C}{t}}$.
	 \subsection{Shamir's Secret Sharing \cite{Sha}}
	 In this sub-section, we discuss the idea of secret sharing. It is a  difficult task to provide demand privacy in multi-access networks just by using privacy keys as every single cache is accessed by multiple users. So, we make use of Shamir's secret sharing technique{\cite{Sha}}. It shows the way to divide data $D$ into $n$ pieces in such a way that $D$ is easily reconstructable from any $k$ out of $n$ pieces, but complete knowledge of any $k-1$ pieces will not give any information about $D$. In order to protect the data, we encrypt it. But, in order to protect the encryption key, we make use of the secret sharing technique as further encryption will not serve the purpose. Generally, the most secured key management schemes keep the key in a single well-guarded location. But, this may make the key inaccessible even when a single misfortune occurs. In the context of multi-access coded  caching, keeping the key in a single cache makes the key accessible to multiple users which destroys the demand privacy constraint. So, we divide the key into multiple pieces called shares and store them across the multiple caches in such a way that only legitimate users can get the key. Let's say $D$ is a number that belongs to a finite field $\mathcal{F}_{p}$, where $p$ is a prime or power of a prime, and $p$ is greater than $n$ . To divide $D$  into multiple pieces $D_1,D_2,...D_n$, pick a random $k-1$ degree polynomial $q(x)=a_{0}+a_{1}x+a_{2}x^{2}....+a_{k-1}x^{k-1}$ where $a_{0}=D$ and $a_{1},a_{2},.......a_{k-1}$ are chosen randomly from $\mathcal{F}_{p}$. Now, evaluate $q(x)$ at $n$ different locations, say at $x=x_{1},x_{2},....x_{n}$ where $x_{i} \in \mathcal{F}_p/0$. These evaluations are called shares. Thus,
	 
	 \begin{center}
	 	$D_{1}=q(x_{1}),D_{2}=q(x_{2}),.....D_{n}=q(x_{n})$
	 \end{center}    
 Let us assume that an opponent who wants to steal $D$ knows $k-1$ shares. As the polynomial is of degree $k-1$ and there are $K$ unknown coefficient values, no information about $D$ can be known to the opponent. Thus, we can protect the data. We make use of this secret sharing scheme to protect the security and privacy keys. We generate $r$ shares where $r$ is the access ratio and store them across different caches in such a way that only intended users will be able to reconstruct them. Now, let's see an example to better understand the secret sharing technique when binary data is considered.

 \begin{exmp}
 	Consider $D =(1\, 0 \, 0 \, 1 \, 0 \, 0 \, 1 \, 1 \, 0 \, 1)$. Our goal here is to generate $4$ shares such that all $4$ are required to reconstruct $D$. Choose a finite field of size $2^{l}$, where $l$ is the smallest positive integer such that $4 < 2^{l} $. So, we choose $\mathcal{F}_{2^{3}} =\{0,1,\alpha,\alpha^{2}, \alpha^{3}, \alpha^{4}, \alpha^{5}, \alpha^{6}\}$. As 3 doesn't divide 10, we append 2 zeros to $D$.  Now, we represent $D$  as a 4-length vector over $\mathcal{F}_8$ where every 3 bits corresponds to one element of $\mathcal{F}_{8}$ as follows : we represent $(0 0 0)$ as $0$, $(0 0 1)$ as $1$, $(0 1 0)$ as $\alpha$, $(0 1 1)$ as $\alpha^{2}$, $(1 0 0)$ as $\alpha^{3}$, $(1 0 1)$ as $\alpha^{4}$, $(1 1 0)$ as $\alpha^{5}$, $(1 1 1)$ as $\alpha^{6}$. So, $D$ becomes $S$= $(\alpha^{3} \, \alpha^{3} \,   \alpha^{5} \,  \alpha^{3})$. Now, we generate shares for each element in $S$ individually. Consider a polynomial of degree 3 with a constant term as the first element of $S$, i.e., $\alpha^{3}$. Let the polynomial be $f(x)= a_{1} x^{3}+a_{2}x^{2}+a_{3}x+\alpha^{3} $, where $\{a_{1},a_{2},a_{3}\} \in \mathcal{F}_{2^{3}}$ . We evaluate $f(x)$ at 4 different locations, say at $x=1, \alpha, \alpha^{2}, \alpha^{3}$. Each evaluation is called a share for that element. So, shares for the element $\alpha^{3}$ are $f(1), f(\alpha), f(\alpha^{2}), f(\alpha^{3})$. As the polynomial is of degree 3, using 4 evaluations, the constant term can be found by interpolation. We follow a similar procedure to generate shares for other elements in $S$. Thus, we can generate 4 shares, ${D^{j} \in \mathcal{F}^{4}_{8}, j \in [4]}$ for $D$. All 4 shares are required to reconstruct $D$. After reconstruction user will discard the zeros that were appended.
 \end{exmp}
	 \subsection{Our Contribution}
	
	We use the idea of key superposition and combine it with a secret sharing technique to achieve demand privacy and security simultaneously for the considered multi-access network. We make use of MDS codes(Maximum Distance Separable) to get better schemes that provide security. The main contributions of this paper are as follows :
	\begin{itemize}
		 
		\item Content security and demand privacy along with linear function retrieval for the network considered in \cite{MKR} has been investigated. Condition on cache memory for security is given.
		\item An achievable SP-LFR scheme that provides security and privacy simultaneously is proposed. The Private-LFR (P-LFR) and LFR schemes can be recovered from our SP-LFR scheme as special cases.
		\item An achievable Secure LFR (S-LFR) scheme, which is a direct extension of the security key scheme for dedicated cache set up in \cite{SeT} is given.
		\item An achievable improved S-LFR scheme that makes use of MDS codes and achieves security with better performance than the S-LFR scheme is proposed.  
		\item In the higher memory region, SP-LFR and P-LFR schemes are shown to have the same performance.
		\item When access ratio, $r=1$, our SP-LFR scheme reduces to  MAN-PDA based SP-LFR scheme in \cite{YaT} and our P-LFR scheme reduces to  MAN-PDA based P-LFR scheme in \cite{YaT}. Security key scheme in \cite{SeT} can also be recovered as a special case from our S-LFR scheme.  
		\item At a memory point, $M=\frac{\binom{C}{r}}{C}$ the improved S-LFR scheme is within a constant multiplicative factor from the optimal rate under uncoded prefetching for $N\geq 2K$.
		\item  At a memory point, $M=\frac{r\binom{C}{r}}{C}$ the  SP-LFR scheme is within a constant multiplicative factor from the optimal rate under uncoded prefetching for $N\geq 2Kr$. 
	\end{itemize}
	\subsection{Paper Organisation}
	Section II introduces problem setting. Section III contains a motivating example and condition on cache memory for security, Section IV contains SP-LFR scheme, Section V contains S-LFR scheme and Section VI contains Improved S-LFR scheme. Section VII contains main results and examples. Section VI concludes the paper. \\

	\textit{Notations:} The set $\{1,2,3,....n\}$ is denoted as $[n]$. |$\chi$| denotes cardinality of the set $\chi$.$\lceil x\rceil$ gives the smallest integer greater than or equal to $x$. A finite field of size $q$ is repreented as $\mathcal{F}_{q}$. For two non-negative integers $n,m$, we have $\binom{n}{m} = \frac{n!}{(n-m)!m!}$, if $n\geq m$, and $\binom{n}{m} =0$ if $n<m$.
	\begin{figure}
		\begin{center}
			\includegraphics[scale=0.8]{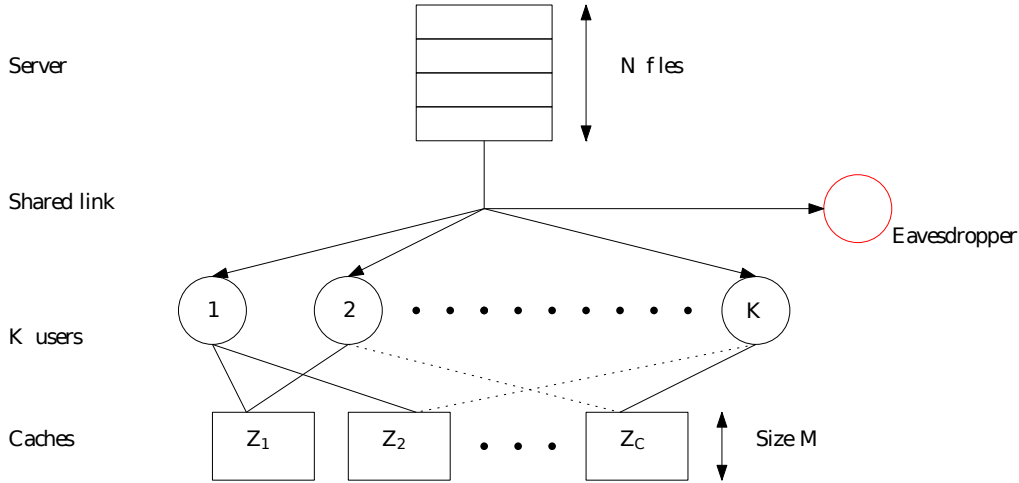}
			\caption {Problem setup for the multi-access setup with $K$ users and $C$ caches}
			\label{fig1}
		\end{center}
	\end{figure}
	\section{Problem Setting}
	Figure \ref{fig1} shows the problem setup. The server $\mathcal{S}$ contains $N$ files denoted as $ W_{1},W_{2},W_{3},....W_{N} $ each of size $F$ bits. It is connected to $K$ users through an error free shared link. Let $\mathcal{K}$ denote the set of $K$ users. Each user is having an access to unique set of $r$ caches out of $C$ caches. Each cache is of size $M$ files. $Z_{c}$ denotes the content of cache $c \in [C]$. 
	Let
	\begin{equation} \label{eq1}
		{\Omega_{r} \triangleq \{g : g \subseteq [C],|g|=r\}}.
	\end{equation}

	\textit{Placement Phase}:
	The server privately generates a random variable $P$ independently and uniformly from a probability space $\mathcal{P}$ and fills the cache of each user by using the cache function
	\begin{equation} \label{eq2}
		{\psi_{c} :\mathcal{P} \times (\mathcal{F}_2)^{NF} \rightarrow (\mathcal{F}_2)^{MF},c\in[C]}. 
	\end{equation}
	The cached content of the cache $c \in [C]$ is denoted by
	\begin{equation} \label{eq3}
		{Z_{c}= \psi_{c}(P,W_{[N]})}. 
	\end{equation}
	Let $ {U_{g}, g \in \Omega_{r} }$ denote the user connected to the set of caches represented by $g$ and $Z_{g}$ represent the set of caches available to the user $U_{g}$.
	\begin{equation} \label{eq4}
		{Z_{g}= \cup_{i\in g} Z_{i}}.
	\end{equation}
	\textit{Delivery Phase}: The demand vector of an user $U_{g}$ is $\textbf{d}_{g} = (d_{{g},1},.....,d_{{g},N}) \in \mathcal{F}^{N}_{2}$. So, the user, $U_{g}$ is interested  in retrieving the  linear combination
	\begin{equation} \label{eq5} 
		{B_{{{g}}} \triangleq d_{{g},1} W_{1}+....+d_{{g},N} W_{N}}.
	\end{equation}
	In order to satisfy the demands of  every user, server transmits 
	\begin{equation} \label{eq6}
		{X=\phi(P,{\textbf{d}_{{\Omega_{r}}}},W_{[N]})}.
	\end{equation}
	where ${\phi :\mathcal{P}\times(\mathcal{F}_2)^{KN} \times(\mathcal{F}_2)^{NF} \rightarrow (\mathcal{F}_2)^{RF}}$ and the quantity $R$ is called the rate of the system. The file library $W_{[N]}$, the randomness $P$ and the demands \{${\textbf{d}_{{g}}} ,\forall g \in \Omega_{r} $ \} are independent. Each user $U_{g}, g \in \Omega_{r}$ must decode its demanded linear combination $B_{{{g}}}$ in (\ref{eq5}), the other non-colluding users should not know anything about the demand for demand privacy and the eaves dropper must not know anything about the contents of library to guarantee security. 
	So, the following are the conditions for correctness, security and privacy: 
	\begin{itemize}
		\item \textit{Correctness}: 
		\begin{equation} \label{eq7}
			H(B_{{{g}}}|X,{\textbf{d}_{{g}}},Z_{{g}}) = 0 ,\quad  \forall g\in\Omega_{r},
		\end{equation}
		\item \textit{Security}: 
		\begin{equation} \label{eq8}
			{I(W_{[N]};X) = 0},
		\end{equation}
		\item \textit{Privacy}:
		\begin{equation} \label{eq9}
			{I(\textbf{d}_{{{\Omega_{r}\backslash g}}};X,\textbf{d}_{{g}},Z_{{g}}) = 0  \quad \forall g \in \Omega_{r}}.
		\end{equation}
	\end{itemize}
	Before we discuss the actual schemes, we present a small example to better understand the schemes in the following section.
	\section{Motivating Example and Condition on Cache Memory for Security }
	\begin{exmp}
		Consider $C=3,r=2,t=1,N=3$. Each file is divided into $3$ sub files. Thus, $W_{i}=\{W_{i,1},W_{i,2},W_{i,3}\} \quad \forall i\in[3]$. Number of users , $K=\binom{C}{r}=\binom{3}{2}=3$. The $3$ users are $\{U_{\{1,2\}}, U_{\{1,3\}}, U_{\{2,3\}}\}$. For providing security, we have to protect each transmission using a security key. So, server generates $\binom{C}{t+r}= \binom{3}{3} = 1$ security key ${V_{\{1,2,3\}}}$ independently and uniformly from $\mathcal{F}_{2}^{F/3}$. To provide demand privacy for every user, server generates $K=3$ random vectors as follows:
	\begin{center}
		$\textbf{P}_{g} \triangleq (p_{g,1},p_{g,2},p_{g,3})^{T} \sim Unif\{\mathcal{F}_{2}^{3}\}, \forall g\in \Omega_{2}$.
	\end{center}
	Using the random vectors, server generates privacy keys as follows :
	\begin{center}
		\begin{itemize}
			\item $T_{\{1,2\},3}= p_{\{1,2\},1}W_{1,3}\oplus p_{\{1,2\},2}W_{2,3} \oplus p_{\{1,2\},3}W_{3,3}$ \\
			\item $T_{\{1,3\},2}= p_{\{1,3\},1}W_{1,3}\oplus p_{\{1,3\},2}W_{2,3} \oplus p_{\{1,3\},3}W_{3,3}$ \\
			\item $T_{\{2,3\},1}= p_{\{2,3\},1}W_{1,1}\oplus p_{\{2,3\},2}W_{2,1} \oplus p_{\{2,3\},3}W_{3,1}$
			
		\end{itemize}
		
	\end{center}
	Our goal here is to provide both security from external eavesdropper and demand privacy simultaneously. So, we make use of the idea of key superposition. We combine both security and privacy keys and place them in cache memory. Let,
	
	\begin{center}
		\begin{itemize}
			\item $D_{\{1,2\},3} = V_{\{1,2,3\}} \oplus T_{\{1,2\},3}$
			\item $D_{\{1,3\},2} = V_{\{1,2,3\}} \oplus T_{\{1,3\},2}$
			\item $D_{\{2,3\},1} = V_{\{1,2,3\}} \oplus T_{\{2,3\},1}$
		\end{itemize}
		
	\end{center}
	As it is a multi-access network, we cannot place the keys directly in the cache. If the keys are placed directly, then more than one user will have access to the key which is dedicated for a particular user there by not satisfying the demand privacy constraint. So, the server makes use of secret sharing technique to generate $2$ shares for $D_{\{1,2\},3},D_{\{1,3\},2},D_{\{2,3\},1}$. Let the $2$ shares of $D_{\{1,2\},3}$ be $D^{1}_{\{1,2\},3}$ and $D^{2}_{\{1,2\},3}$. Both shares are required to reconstruct $D_{\{1,2\},3}$. We place them in such a way that only user $U_{\{1,2\}}$ can have access to both shares. Thus, other users cannot know anything about $D_{\{1,2\},3}$. Similarly $D^{1}_{\{1,3\},2}, D^{2}_{\{1,3\},2}$ and $D^{1}_{\{2,3\},1}, D^{2}_{\{2,3\},1}$ are shares for $D_{\{1,3\},2}$ and $D_{\{2,3\},1}$ respectively. The placement is as follows :
	\begin{itemize}
		\item $Z_{1} = \{W_{{1,1}}, W_{{2,1}}, W_{{3,1}}\} \cup \{D^{1}_{\{1,2\},3}, D^{1}_{\{1,3\},2} \} $
		\item $Z_{2} = \{W_{{1,2}}, W_{{2,2}}, W_{{3,2}}\} \cup \{D^{2}_{\{1,2\},3}, D^{2}_{\{2,3\},1} \} $
		\item $Z_{3} = \{W_{{1,3}}, W_{{2,3}}, W_{{3,3}}\} \cup \{D^{3}_{\{1,3\},2}, D^{3}_{\{2,3\},1} \} $ 
	\end{itemize} 
	Each cache has $3$ sub files and 2 shares. Size of each sub file is $\frac{1}{3}^{rd}$ of a file and size of each share is same as size of the sub file. Thus , total cache size is $\frac{5}{3}$ files. \\
	Let the user $U_{\{1,2\}}$ be requesting $W_{1}$, $U_{\{1,3\}}$ be requesting $W_{2}$, $U_{\{2,3\}}$ be requesting $W_{3}$. In order to satisfy the requests, server transmits the following :
	\begin{itemize}
		\item $X= V_{\{1,2,3\}} \oplus W_{1,3} \oplus T_{\{1,2\},3} \oplus W_{2,2} \oplus T_{\{1,3\},2} \oplus W_{3,1} \oplus T_{\{2,3\},1}$
		\item $\textbf{q}_{g} = \textbf{P}_{g} + \textbf{d}_{{g}} \quad \forall g \in \Omega_{2} $ 
	\end{itemize}
	Now, consider a user $U_{\{1,2\}}$. It can reconstruct $D_{\{1,2\},3}$, as it has access to $D^{1}_{\{1,2\},3}$ and $D^{2}_{\{1,2\},3}$. It is clear that no other user can reconstruct  $D_{\{1,2\},3}$ as both shares are required for the reconstruction. $U_{\{1,2\}}$ can get $W_{2,2} \oplus T_{\{1,3\},2}= q_{\{1,3\},1}W_{{1,2}}\oplus q_{\{1,3\},2}W_{{2,2}} \oplus q_{\{1,3\},3}W_{{3,2}}$ as $\textbf{q}_{\{1,3\}}=(q_{\{1,3\},1},q_{\{1,3\},2},q_{\{1,3\},3})$ and $W_{1,2}, W_{2,2}, W_{3,2}$ are accessible to it. Similarly, it can get $W_{3,1} \oplus T_{\{2,3\},1}$. Thus, it can get the required sub file $W_{1,3}$. Similarly, the other users can also get the files they wanted. The transmission is protected by a security key. So, it is secure from any external eavesdropper. As, $\textbf{q}_{g} \forall g \in \Omega_{2}$ is uniformly distributed over $\mathcal{F}^{3}_{2}$, demand privacy is satisfied.  Rate achieved is $\frac{1}{3}$. The above scheme achieves both demand privacy and security simultaneously. If the requirement is only security, then we present $2$ different schemes. Firstly, we apply S-LFR scheme. Server generates only security keys and placement is done as follows :
	\begin{itemize}
		\item  $Z_{1} = \{W_{{1,1}}, W_{{2,1}}, W_{{3,1}}\} \cup \{V_{\{1,2,3\}}\}$
		\item $Z_{2} = \{W_{{1,2}}, W_{{2,2}}, W_{{3,2}}\} \cup \{V_{\{1,2,3\}}\}$
		\item $Z_{3} = \{W_{{1,3}}, W_{{2,3}}, W_{{3,3}}\} \cup
		\{V_{\{1,2,3\}}\}$
	\end{itemize}
	The cache memory occupied is $\frac{4}{3}$. The transmission is as follows:
	\begin{center}
		$X= V_{\{1,2,3\}} \oplus W_{1,3}  \oplus W_{2,2} \oplus W_{3,1}$ 
	\end{center}
	The above transmission is protected by a security key. Consider a user $U_{\{1,2\}}$, it has $W_{\{2,2\}}, W_{\{3,1\}}$ in its caches and it also has the security key. Thus, it can get its demanded sub file $W_{1,3}$. Rate is $\frac{1}{3}$. Now, let us see improved S-LFR scheme whose performance is better than the S-LFR scheme.
	By changing the way we place the security key, we can have the advantage of saving the cache memory. Consider the security key $V_{\{1,2,3\}}$ and do the following :
	\begin{itemize}
		\item Divide the security key in to $2$ sub-keys. $V_{\{1,2,3\}}=\{V_{\{1,2,3\},1},V_{\{1,2,3\},2}\}$.
		\item Encode using the following (3,2) MDS code
		\begin{center}
			
			$\textbf{A}= \begin{bmatrix}
			1 & 0 & 1 \\
			0 & 1 & 1
			
		\end{bmatrix}$
		\end{center} 
		\item Now, the three coded sub-keys are:
		\begin{itemize}
		\item $\tilde{V}_{\{1,2,3\},1}=V_{\{1,2,3\},1}$
		\item $\tilde{V}_{\{1,2,3\},2}=V_{\{1,2,3\},2}$
		\item $\tilde{V}_{\{1,2,3\},3}=V_{\{1,2,3\},1} \oplus V_{\{1,2,3\},2} $
		\end{itemize}
		\end{itemize}
		Now, the placement is as follows:
		\begin{itemize}
		\item  $Z_{1} = \{W_{{1,1}}, W_{{2,1}}, W_{{3,1}}\} \cup \{\tilde{V}_{\{1,2,3\},1}\}$
		\item $Z_{2} = \{W_{{1,2}}, W_{{2,2}}, W_{{3,2}}\} \cup \{\tilde{V}_{\{1,2,3\},2}\}$
		\item $Z_{3} = \{W_{{1,3}}, W_{{2,3}}, W_{{3,3}}\} \cup
		\{\tilde{V}_{\{1,2,3\},3}\}$
		\end{itemize}
		
		By the above placement, it is clear that each user can have access to $2$ coded sub-keys which are enough to reconstruct the entire security key. The size of coded sub-key is same as the size of sub-key which is $\frac{1}{6}^{th}$ of a file. So, the cache memory is $\frac{7}{6}$ which is less than the memory occupied by SLFR scheme.  The transmission is as follows :
		\begin{center}
			$X= V_{\{1,2,3\}} \oplus W_{1,3}  \oplus W_{2,2} \oplus W_{3,1}$ 
		\end{center}
		The above transmission is protected by a security key. Consider a user $U_{\{1,2\}}$, it has $W_{2,2}, W_{3,1}$ in its caches and it can also get the security key as it has access to $2$ coded sub-keys from its caches. Thus, it can get its demanded sub file $W_{1,3}$. Rate is $\frac{1}{3}$.
	\end{exmp}
		
		Before we present the actual scheme, we show that for secure delivery, the cache memory $M\geq\frac{\binom{C}{r}}{C}$.
		\begin{prop}
			For a secure multi-access coded caching scheme with distinct demands, the cache memory $M\geq \frac{\binom{C}{r}}{C}$.
		\end{prop}
	\textit{Proof}: Consider a multi-access network consisting of  a server that contain $N$ files which is connected to $K$ users through an error free shared link. Each user is having an access to a unique set of $r$ caches out of $C$ caches.
		We prove the condition considering single file retrieval. It is valid for LFR as well. For the considered network, total number of users is $K=\binom{C}{r}$. We consider the worst case scenario where each user demands a file which is not demanded by any other user. Let the demand vector be $\textbf{d}=[d_{1}, d_{2}....,d_{K}]$. We can write
		\begin{equation}\label{eq10}
		H(W_{d_{1}},W_{d_{2}},.....W_{d_{K}}|X,Z_{1},Z_{2},...Z_{C})=0
		\end{equation}
		\begin{equation}\label{eq11}
		I(W_{d_{1}}, W_{d_{2}},.....W_{d_{K}}; X)=0
		\end{equation}
		The above equations (\ref{eq10}), (\ref{eq11}) corresponds to decodability and security conditions respectively. We have,
		\begin{subequations}
			\begin{align}
			\binom{C}{r}F &= H(W_{d_{1}}, W_{d_{2}},.....W_{d_{K}}) \\
			&=I(W_{d_{1}},W_{d_{2}},.....W_{d_{K}};X,Z_{1},Z_{2},...Z_{C})  +H(W_{d_{1}}, W_{d_{2}},.....W_{d_{K}}|X,Z_{1},Z_{2},...Z_{C}) \\
			&= I(W_{d_{1}}, W_{d_{2}},.....W_{d_{K}};X,Z_{1},Z_{2},...Z_{C}) \\
			&=I(W_{d_{1}}, W_{d_{2}},.....W_{d_{K}};X)+I(W_{d_{1}}, W_{d_{2}},.....W_{d_{K}};Z_{1},Z_{2},...Z_{C}|X) \\
			&=I(W_{d_{1}}, W_{d_{2}},.....W_{d_{K}};Z_{1},Z_{2},...Z_{C}|X) \\
			&\leq H(Z_{1},Z_{2},...Z_{C}) \\
			&\leq \sum_{c=1}^{C} H(Z_{c}) \\
			&\leq CMF 
			\end{align}
		\end{subequations}
		This implies $M \geq \frac{\binom{C}{r}}{C}$.
		
	\section{SP-LFR SCHEME}
	For $t \in [0: C-r] $, each file is divided in to $\binom{C}{t}$ sub files as follows :
	\begin{equation} \label{eq13}
		{W_{i}=\{W_{i,\mathcal{T}} : \mathcal{T} \in \Omega_{t} \}, \forall i\in[N]}.	
	\end{equation}
	\textit{Placement Phase:}  The server generates $ \binom{C}{t+r}$ security keys and $ \binom{C}{r} \binom{C-r}{t}$ privacy keys as follows:
	\begin{itemize}
		\item The server generates $ \binom{C}{t+r}$ security keys ${\{V_{S} : S \in \Omega_{t+r}\}}$ independently and uniformly from $\mathcal{F}_{2}^{F/\binom{C}{t}}$.
		\vspace{0.5cm}
		\item The $ \binom{C}{r} \binom{C-r}{t}$ privacy keys are generated as follows. The server first generates $\binom{C}{r}$ i.i.d. random vectors $\{\textbf{P}_{g} : g \in \Omega_{r}\}$ from $\mathcal{F}^{N}_{2}$ , i.e.
		\begin{equation} \label{eq14}
			\textbf{P}_{g} \triangleq (p_{g,1},...p_{g,N})^{T} \sim Unif\{\mathcal{F}_{2}^{N}\}, \forall g\in \Omega_{r}.
		\end{equation}

		Then the $ \binom{C}{r} \binom{C-r}{t}$ privacy keys, denoted by $\{T_{g,\mathcal{T}}; g \in \Omega_{r} , \mathcal{T} \in \Omega_{t},g \cap \mathcal{T}= \emptyset\}$ are generated as :
		\begin{equation} \label{eq15}
			{T_{g,\mathcal{T}}\triangleq \underset{n\in[N]}{\sum} p_{g,n}.W_{n,\mathcal{T}}}.
		\end{equation} 
	\end{itemize}    
	Let's define $D_{g,\mathcal{T}}$ as follows:
	\begin{equation} \label{eq16}
		D_{g,\mathcal{T}} \triangleq T_{g,\mathcal{T}} \oplus V_{g \cup \mathcal{T}} ,\quad  g \in \Omega_{r} , \mathcal{T} \in \Omega_{t},g \cap \mathcal{T}= \emptyset .
	\end{equation}

\begin{figure}
		\begin{center}

			\includegraphics[scale=0.8]{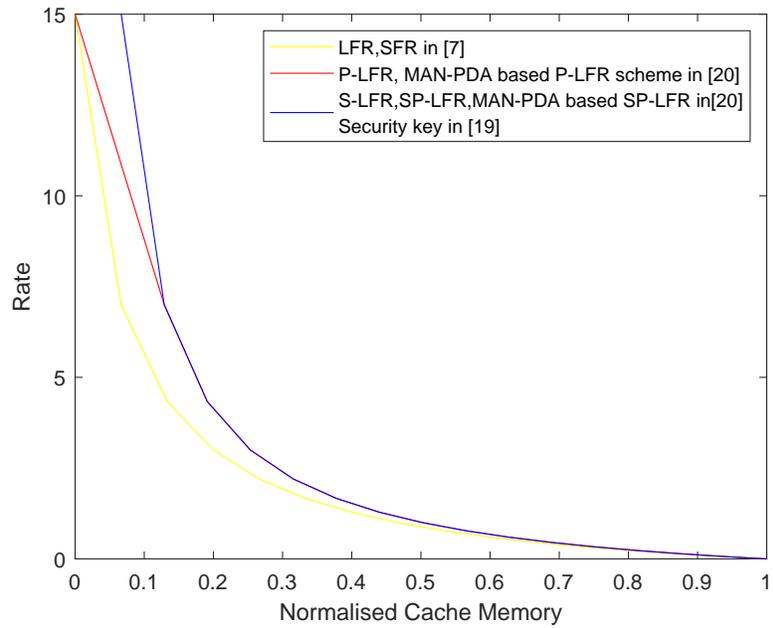}

			\caption {Performance of the multi-access network for different schemes when $C=15, r=1, K=N=15$}
			\label{fig2}
		\end{center}
	\end{figure}

\begin{figure}
		\begin{center}

			\includegraphics[scale=0.8]{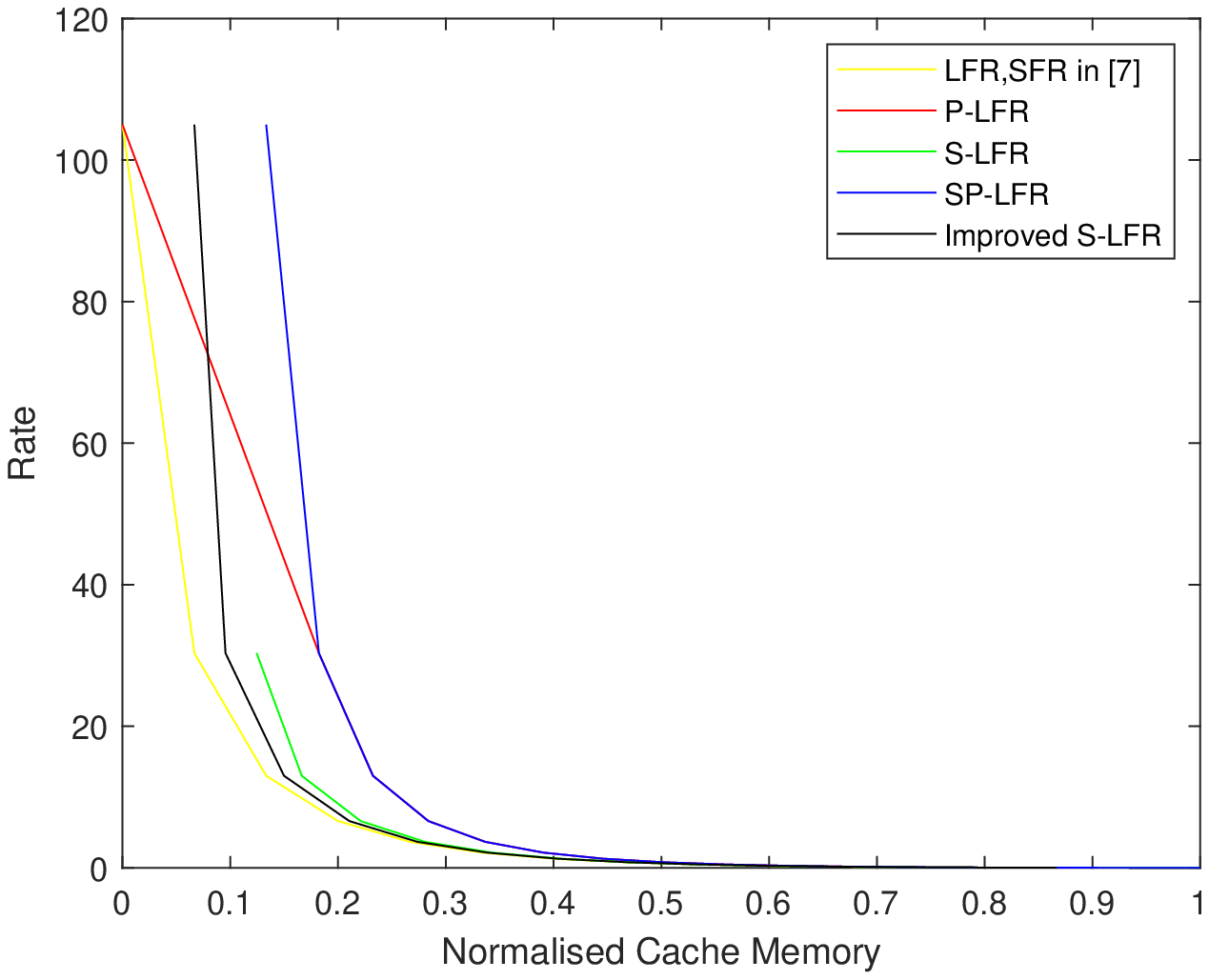}

			\caption {Performance of the  multi-access network for different schemes when $C=15, r=2, K=N=105$}
			\label{fig3}
		\end{center}
	\end{figure}

\begin{figure}
		\begin{center}

			\includegraphics[scale=0.8]{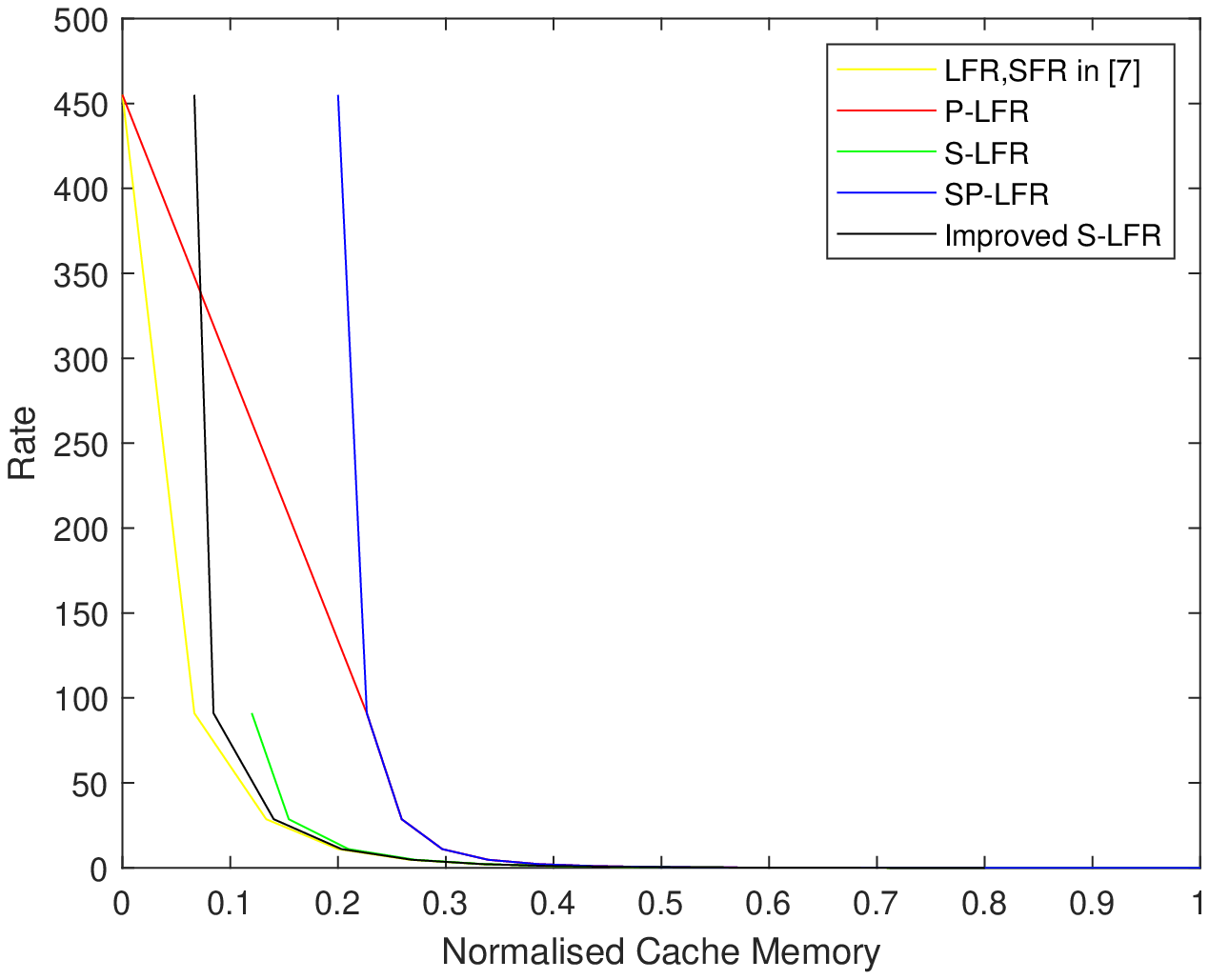}

			\caption {Performance of the multi-access network for different schemes when $C=15, r=3, K=N=455$}
			\label{fig4}
		\end{center}
	\end{figure}

Now, the server uses Shamir's secret sharing technique \cite{Sha} on $D_{g,\mathcal{T}}$  to generate $r$ random vectors called shares such that $D_{g,\mathcal{T}}$  could be re-constructed from $r$ shares and we cannot get any information about $D_{g,\mathcal{T}}$ by any $r - 1$ shares or less. Each share has $\left\lceil \frac{F}{\binom{C}{t}l} \right\rceil$ elements where each element belongs to $\mathcal{F}_{2^{l}}$ ($l$ is the smallest positive integer such that $r < 2^{l} $).
	Let $g(j)$ represent the $j^{th}$ element in the set $g$. Then, the content of each cache is given by:
	\begin{multline} \label{eq17}
		Z_{c} = \{W_{i,\mathcal{T}} : c\in\mathcal{T} , \mathcal{T} \in \Omega_{t} \quad \forall \quad i \in [N] \} \\ 
		\cup  \quad  \{D^{j}_{g,\mathcal{T}} : c \in g ,g \in \Omega_{r} ,        \mathcal{T} \in \Omega_{t}, g \cap \mathcal{T}= \emptyset, c=g(j)\} .
	\end{multline}
	\textit{Cache Memory}: Let's call the part of  cache memory that is allocated for storing the sub files as data memory and the other part for storing the keys as key memory. $\binom{C-1}{t-1}$ sub files of each file are placed in every cache. So, the size of data memory is equal to $\frac{tN}{C}=M^{D}_{SP}$ . As the size of each share is $\left\lceil \frac{F}{\binom{C}{t}l} \right\rceil\frac{l}{F}$, the size of key memory is equal to $\binom{C-r}{t} \binom{C-1}{r-1} \left\lceil \frac{F}{l\binom{C}{t}} \right\rceil\frac{l}{F}= M^{K}_{SP}$. So , the size of each cache is as follows:
	\begin{equation} \label{eq18}
		 M_{SP}=\frac{Nt}{C}+\binom{C-r}{t} \binom{C-1}{r-1} \left\lceil \frac{F}{l\binom{C}{t}} \right\rceil\frac{l}{F}
	\end{equation}
	\textit{Delivery Phase}: In order to satisfy the demands of each user as in (\ref{eq5}), the server transmits $X=[Y_{S}, \textbf{q}_{g}] \quad \forall S \in \Omega_{t+r}, g \in \Omega_{r}  $ , where 
	\begin{equation} \label{eq19}
		{Y_{S} = V_{S} \oplus {\underset{\substack{g\subset S\\|g|=r}}{\bigoplus} B_{{{g}},S\backslash g} \oplus T_{g,S\backslash g}} \quad \forall S \in \Omega_{t+r}} ,
	\end{equation}
	\begin{equation} \label{eq20}
		{	\textbf{q}_{g} = \textbf{d}_{{g}} \oplus \textbf{P}_{g} \quad \forall g \in \Omega_{r} }.
	\end{equation}
	\textit{Correctness}: Consider an user $U_{h} , h \subset S, S \in \Omega_{t+r}$ and the transmission $Y_{S}$ can be written as  
	\begin{equation} \label{eq21}
		Y_{S} = V_{S} \oplus B_{{{h}},S\backslash h} \oplus T_{h,S\backslash h} {\underset{\substack{g\subset S\\|g|=r \\ g\neq h} }{\bigoplus} B_{{{g}},S\backslash g} \oplus T_{g,S\backslash g}}.
	\end{equation}
	From the placement, the user $U_{h}$ can get $D_{h,S\backslash h} = T_{h,S\backslash h} \oplus V_{S}$. Now, consider the term 
		$B_{{{g}},S\backslash g} \oplus T_{g,S\backslash g}$,
	and it can be written in the following way
	\begin{equation} \label{eq23}
		(p_{g,1} \oplus d_{{g},1})W_{1,S\backslash g} \oplus ..........\oplus (p_{g,1} \oplus d_{{g},N})W_{N,S\backslash g}.
	\end{equation}
	The coefficients of $W_{i,S\backslash g},  \forall i \in [N]$ are known to the user as $\textbf{q}_{g} , \forall g \in \Omega_{r} $ are sent by the server and the user $U_{h}$ has access to the sub files $W_{i,S\backslash g}, g \subset S,|g|=r, g\neq h  \forall i \in [N]$. So, it can calculate the entire term. Thus, the user gets $B_{{{h}},S\backslash h}$. Similarly, from all the transmissions corresponding to  such $t+r$ sized subsets $S$ , with $h\subset S$, the user $U_{h}$ gets the missing sub files of its demand. Since this is true for every such subset $S$, any user $U_{g},  g\subset C,|g|=r$ can recover the missing sub files of its demanded file. \\
	\textit{Privacy}: We now prove that the above scheme satisfies the privacy condition in (\ref{eq9}).
	\begin{subequations}
		\begin{align}
			I(\textbf{d}_{{\Omega_{r}\backslash g}};X,\textbf{d}_{{g}},Z_{{g}}) 
			&\leq I(\textbf{d}_{{\Omega_{r}\backslash g}};X,\textbf{d}_{{g}},Z_{{g}},W_{[N]}) \\
			&=I(\textbf{d}_{{\Omega_{r}\backslash g}};\textbf{q}_{{\Omega_{r}}},Y,\textbf{d}_{{g}},Z_{{g}},W_{[N]}) \\
			&\leq I(\textbf{d}_{{\Omega_{r}\backslash g}};\textbf{q}_{{\Omega_{r}}},V_{\Omega_{t+r}},Y,\textbf{d}_{{g}},Z_{{g}},W_{[N]}) \\
			&=I(\textbf{d}_{{\Omega_{r}\backslash g}};\textbf{q}_{{\Omega_{r}}},V_{\Omega_{t+r}},\textbf{d}_{{g}},Z_{{g}},W_{[N]}) \\
			&\leq I(\textbf{d}_{{\Omega_{r}\backslash g}};\textbf{q}_{{\Omega_{r}}},a_{[r-1]},V_{\Omega_{t+r}},\textbf{P}_{g},\textbf{d}_{{g}},Z_{{g}},W_{[N]}) \\ 	
			&= I(\textbf{d}_{{\Omega_{r}\backslash g}};\textbf{q}_{{\Omega_{r}\backslash g}},V_{\Omega_{t+r}},\textbf{P}_{g},a_{[r-1]},\textbf{d}_{{g}},W_{[N]}) \\ 		
			&= I(\textbf{d}_{{\Omega_{r}\backslash g}};V_{\Omega_{t+r}},\textbf{P}_{g},a_{[r-1]},\textbf{d}_{{g}},W_{[N]}) + I(\textbf{d}_{{\Omega_{r}\backslash g}};\textbf{q}_{{\Omega_{r}\backslash g}} | V_{\Omega_{t+r}},\textbf{P}_{g},a_{[r-1]},\textbf{d}_{{g}},W_{[N]})  \\ 
			&=I(\textbf{d}_{{\Omega_{r}\backslash g}};\textbf{q}_{{\Omega_{r}\backslash g}} | V_{\Omega_{t+r}},\textbf{P}_{g},a_{[r-1]},\textbf{d}_{{g}},W_{[N]})  \\
			& =H(\textbf{q}_{{\Omega_{r}\backslash g}} | V_{\Omega_{t+r}},\textbf{P}_{g},a_{[r-1]},\textbf{d}_{{g}},W_{[N]}) - H(\textbf{q}_{{\Omega_{r}\backslash g}} |  V_{\Omega_{t+r}},\textbf{P}_{g},\textbf{d}_{\Omega_{r}}, a_{[r-1]}, W_{[N]})  \\
			& =H(\textbf{q}_{{\Omega_{r}\backslash g}}) - H(\textbf{P}_{{\Omega_{r}\backslash g}} |  V_{\Omega_{t+r}},\textbf{P}_{g},\textbf{d}_{\Omega_{r}},a_{[r-1]},W_{[N]})  \\
			&=H(\textbf{q}_{{\Omega_{r}\backslash g}}) - H(\textbf{P}_{{\Omega_{r}\backslash g}}) \\
			&=0,
		\end{align}
	\end{subequations}
	where (23d) comes from that $Y$ is determined by $\textbf{q}_{{\Omega_{r}}} , V_{\Omega_{t+r}} , W_{[N]}$, (23f) comes from that $Z_{{g}}$ is determined by $ \textbf{P}_{g} , W_{[N]}, V_{\Omega_{t+r}},a_{[r-1]}$, (23j) comes from that $\textbf{q}_{{\Omega_{r}\backslash g}}$ are independent of $(V_{\Omega_{t+r}},\textbf{P}_{g}, a_{[r-1]}, \textbf{d}_{{g}},W_{[N]})$ and $\textbf{q}_{{\Omega_{r}\backslash g}}$ and $\textbf{P}_{{\Omega_{r}\backslash g}}$ determines each other given $\textbf{d}_{{\Omega_{r}\backslash g}}$, (23l) comes from that $\textbf{q}_{{\Omega_{r}\backslash g}}$ and $\textbf{P}_{{\Omega_{r}\backslash g}}$ are independent and uniformly distributed over $\mathcal{F}^{N}_{2}$. \\
	\textit{Security}: We prove that the above scheme satisfies the security condition in (\ref{eq8})
	\begin{subequations}
		\begin{align}
			I(W_{[N]}; X) &= I(W_{[N]}; \textbf{q}_{\Omega_{r}}, Y) \\
			&= I(W_{[N]}; \textbf{q}_{\Omega_{r}}) + I(W_{[N]}; Y |\textbf{q}_{\Omega_{r}})  \\
			&=0,
		\end{align}
	\end{subequations}
	where (24c) comes from that $\textbf{q}_{\Omega_{r}}$ is independent of $W_{[N]}$ and $Y$ is independent of $W_{[N]}$ , $\textbf{q}_{\Omega_{r}} $.
	\begin{rem}
		If the security keys are removed, the SP-LFR scheme reduces to a P-LFR scheme. But, additionally the memory-rate point $(0,N)$ is achievable for P-LFR scheme due to which it has better rate in the lower memory region which can be seen from the Figure \ref{fig3} and Figure \ref{fig4} where we considered the value of F to be such that l(=2) divides $\frac{F}{\binom{C}{t}} \; \forall \; t \in [0:C-r]$.  If both the security and privacy keys are removed, the SP-LFR scheme reduces to  LFR scheme and LFR scheme reduces to SFR scheme in \cite{MKR} when users demand a single file. For $r=1$,  SP-LFR scheme proposed above reduces to MAN-PDA based SP-LFR scheme in \cite{YaT}, P-LFR scheme reduces to MAN-PDA based P-LFR scheme in \cite{YaT} which can be seen in Figure \ref{fig2}. It must be noted that multi-access network reduces to dedicated cache network when $r$=1 and there is no need of generating shares.
	\end{rem}
\begin{figure}
	\begin{center}

		\includegraphics[scale=0.8]{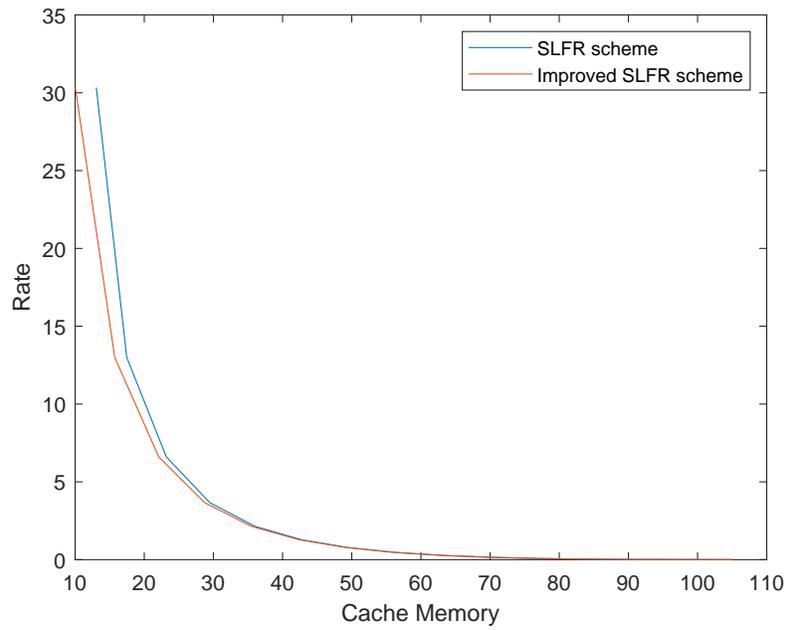}

		\caption {Performance of S-LFR and Improved S-LFR schemes when $C=15, r=2, K=N=105$}
		\label{fig5}
	\end{center}
\end{figure}
	\section{S-LFR scheme} 
		
	For $t\in[0:C-r] $, each file is divide as in (\ref{eq13}). \\
	\textit{Placement Phase} : The server generates $ \binom{C}{t+r}$ security keys ${\{V_{S} : S \in \Omega_{t+r}\}}$ independently and uniformly from $\mathcal{F}_{2}^{F/\binom{C}{t}}$. The content of each cache becomes
	\begin{equation}
		Z_{c} = \{W_{i,\mathcal{T}} : c\in\mathcal{T} , \mathcal{T} \in \Omega_{t} \quad \forall \quad i \in [N] \} 
		\cup   \{V_{S} : c \in S ,S \in \Omega_{t+r},\}.
	\end{equation}
	From the above placement, it is clear that the key memory size is $\frac{ \binom{C-1}{t+r-1}}{\binom{C}{t}}$ and data memory size is $\frac{Nt}{C}$. So, the total memory size is :
	\begin{center}
		$M_{S}=\left(\frac{Nt}{C}+\frac{ \binom{C-1}{t+r-1}}{\binom{C}{t}}\right)$.
	\end{center}
	\textit{Delivery Phase}: Each user requests the demands as in (\ref{eq5}). So, in order to satisfy the demands  server transmits
	\begin{equation}
		{Y_{S} = V_{S} \oplus {\underset{\substack{g\subset S\\|g|=r}}{\bigoplus} B_{{{g}},S\backslash g} } \quad \forall S \in \Omega_{t+r}}.
	\end{equation}
	\textit{Correctness}: 
	A user $U_{g}$ has access to sub file $W_{i,\mathcal{T}}$ ,$\forall$ $i\in [N]$ when $g\cap \mathcal{T}$ $\neq \emptyset $ and to the security key $V_{S}$ when $g \cap S \neq \emptyset$. So, the user $U_{g}$ can recover $B_{{{g}},\mathcal{T}}$ from its own cache when $g\cap \mathcal{T}$ $\neq \emptyset $. Now, consider the transmission
	\begin{center}
		${\underset{\substack{g\subset S\\|g|=r}}{\bigoplus} B_{{{g}},S\backslash g}}$.                                  
	\end{center}
	corresponding to the subset of caches, $S\subset C$ with $|S|=t+r$. The user $U_{g}$ has access to $B_{{{g}},S\backslash h}$ for any other subset $h \subset \{S:h\neq g\}$,$|h|=r$ since, $\{S\backslash h\}\cap g \neq \emptyset $. Hence it can retrieve the sub file $B_{{{g}},S\backslash g}$ from the transmission corresponding to $S$. Similarly, from all the transmissions corresponding to  such $t+r$ sized subsets $S$, where $g\subset S$, user $U_{g}$ gets the missing sub files of its demand. Since this is true for every such subset $S$, any user  $U_{h}, h\subset C,|h|=r$ can recover the missing sub files of its demanded file.\\
	\textit{Security}: The security is guaranteed since each transmitted signal is accompanied by a security key.
	\begin{rem}
		If security keys are removed, the above S-LFR scheme reduces to LFR scheme which has the same performance as SFR scheme in \cite{MKR} as shown in Figure \ref{fig2}, Figure \ref{fig3}, Figure \ref{fig4} where we considered the value of F to be such that l(=2) divides $\frac{F}{\binom{C}{t}} \; \forall \; t \in [0:C-r]$ . When $r=1$, if each user demands  one individual file instead of linear combination of files, the S-LFR scheme reduces to security key scheme in \cite{SeT}.  
	\end{rem}
\section{IMPROVED S- LFR SCHEME}
For $t \in [0: C-r] $, each file is divided in to $\binom{C}{t}$ sub files as follows :
\begin{equation} \label{eq28}
{W_{i}=\{W_{i,\mathcal{T}} : \mathcal{T} \in \Omega_{t} \}, \forall i\in[N]}.	
\end{equation}
\textit{Placement Phase} : The server generates $ \binom{C}{t+r}$ security keys ${\{V_{S} : S \in \Omega_{t+r}\}}$ independently and uniformly from $\mathcal{F}_{2}^{F/\binom{C}{t}}$. The following is done for the generated key:
\begin{itemize}
	\item Divide each key  in to $r$ sub-keys. If $r$ doesn't divide $\frac{F}{\binom{C}{t}}$, append sufficient number of zeros. So, the size of each sub-key is $\left\lceil\frac{F}{r\binom{C}{t}} \right\rceil \frac{1}{F}$ files.
	\item Encode them using a ( t+r, r) MDS (Maximum Distance Separable) code. The coded sub-keys of a security key $V_{S}, S \in \Omega_{t+r}$ are represented as
	$\{\tilde {V}_{S,j} : j \in S, |j|=1 \}$. The size of each coded sub-key is same as the size of a sub-key.
\end{itemize}
The placement is done as follows:
\begin{equation}
Z_{c} = \{W_{i,\mathcal{T}} : c\in\mathcal{T} , \mathcal{T} \in \Omega_{t} \quad \forall \quad i \in [N] \} \\
\cup   \{\tilde{V}_{S,j} : c \in S ,S \in \Omega_{t+r},j=c  \}.
\end{equation}
The above placement ensures the users to get the security key it requires from the caches that are accessible to it. Consider an user  $U_{g}$, $g \in \Omega_{r}$ and a security key $V_{S}, S \in \Omega_{t+r}, g \in S$. The user gets $r$ coded sub-keys of the security key. As the keys are encoded using a (t+r, r) MDS code, $r$ coded sub-keys are enough to retrieve the entire key. \\
\textit{Cache Memory}: Let's call the part of  cache memory that is allocated for storing the sub files as data memory and the other part for storing the keys as key memory. $\binom{C-1}{t-1}$ sub files of each file are placed in every cache. So, the size of data memory is equal to $\frac{tN}{C}=M^{D}_{SP}$. The coded sub-key has the same size as the size of sub-keys. Each cache has a single coded sub-key from $\binom{C-1}{t+r-1}$ security keys. Thus, the overall size of each cache is:
\begin{center}
	$M_{IS}=\left(\frac{Nt}{C}+{ \binom{C-1}{t+r-1}}\lceil \frac{F}{r\binom{C}{t}} \rceil \frac{1}{F} \right)$.
\end{center}
\textit{Delivery Phase}: Each user requests the demands as in (\ref{eq5}). So, in order to satisfy the demands  server transmits
\begin{equation}
{Y_{S} = V_{S} \oplus {\underset{\substack{g\subset S\\|g|=r}}{\bigoplus} B_{{{g}},S\backslash g} } \quad \forall S \in \Omega_{t+r}}.
\end{equation}
\textit{Correctness}: 
An user $U_{g}$ has access to the sub file $W_{i,\mathcal{T}}$ ,$\forall$ $i\in [N]$ when $g\cap \mathcal{T}$ $\neq \emptyset $ and to the security key $V_{S}$ when $g \cap S \neq \emptyset$. So, the user $U_{g}$ can recover $B_{{{g}},\mathcal{T}}$ from its own cache when $g\cap \mathcal{T}$ $\neq \emptyset $. Now, consider the transmission
\begin{center}
	${\underset{\substack{g\subset S\\|g|=r}}{\bigoplus} B_{{{g}},S\backslash g}}$.                                  
\end{center}
corresponding to the subset of caches, $S\subset C$ with $|S|=t+r$. The user $U_{g}$ has access to $B_{{{g}},S\backslash h}$ for any other subset $h \subset \{S:h\neq g\}$,$|h|=r$ since, $\{S\backslash h\}\cap g \neq \emptyset $. Hence it can retrieve the sub file $B_{{{g}},S\backslash g}$ from the transmission corresponding to $S$. Similarly, from all the transmissions corresponding to  such $t+r$ sized subsets $S$ , with $g\subset S$, user $U_{g}$ gets the missing sub files of its demand. Since this is true for every such subset $S$, any user  $U_{h}, h\subset C,|h|=r$ can recover the missing sub files of its demanded file.\\
\textit{Security}: The security is guaranteed since each transmitted signal is accompanied by a security key.
\begin{rem}
	The performance comparision of Improved S-LFR scheme with S-LFR scheme is shown in Figure \ref{fig5}. It is clear that the Improved S-LFR scheme performs better.
\end{rem}

	\section{MAIN RESULTS}
	\begin{thm}
		For a shared link Multi-access network with $C$ caches, access degree r, $K$ users, $N$ files there exists a SP-LFR scheme that achieves the lower convex envelope of the memory-rate pairs in $\{(M_{SP},R_{SP}): t \in [0:C-r]\} \cup \{(N,0)\}$ where 
		\begin{equation*}
			{(M_{SP},R_{SP})=\left(\frac{Nt}{C}+\binom{C-r}{t} \binom{C-1}{r-1} \left\lceil \frac{F}{l\binom{C}{t}} \right\rceil\frac{l}{F},\frac{\binom{C}{t+r}}{\binom{C}{t}}\right)} ,
		\end{equation*}
	where $l$ is the smallest positive integer such that $r< 2^{l}$	with subpacketization $\binom{C}{t}$
	\end{thm}
	
	{\it Proof:}
		In the proposed SP-LFR scheme above, each file is divided in to $\binom{C}{t}$ sub files. So,  the size of each sub file is $\frac{1}{\binom{C}{t}}$. Each transmission is also having a size of $\frac{1}{\binom{C}{t}}$. For every $t+r$ sized subset of $C$, there is a transmission. Since the lengths of $q_{g } , \forall g \in \Omega_{r} $ are negligible compared to the file size, the total rate $R=\frac{\binom{C}{t+r}}{\binom{C}{t}}$.

	\begin{thm}
		For a shared link Multi-access network with $C$ caches, access degree r, $K(=\binom{C}{r}) $ users, $N $ files there exists a P-LFR scheme that achieves the lower convex envelope of the memory-rate pairs in $\{0,N\} \cup \{(M_{SP},R_{SP}): t \in [0:C-r]\} \cup \{(N,0)\}$ where 
		\begin{equation*}
			{(M_{P},R_{P})=\left(\frac{Nt}{C}+\binom{C-r}{t} \binom{C-1}{r-1} \left\lceil \frac{F}{l\binom{C}{t}} \right\rceil\frac{l}{F},\frac{\binom{C}{t+r}}{\binom{C}{t}}\right)} ,
		\end{equation*}
	where $l$ is the smallest positive integer such that $r< 2^{l}$	with subpacketization $\binom{C}{t}$.
	\end{thm}
	
	{\it Proof:}
		In the proposed SP-LFR scheme above, if we do not consider  the security key , it becomes P-LFR Scheme. So, in P-LFR scheme 
		\begin{equation}
			D_{g,\mathcal{T}} \triangleq T_{g,\mathcal{T}}  ,\quad  g \in \Omega_{r} , \mathcal{T} \in \Omega_{t},g \cap \mathcal{T}= \emptyset .
		\end{equation}
		Addiitonally $(0,N)$ point is achievable if only demand privacy is considered. The memory size and rate equations for $t \in [0:C-r]$ remains same as SP-LFR scheme.
	
	\begin{thm}
		For a shared link Multi-access network with $C$ caches, access degree r, $K$ users, $N$ files there exists a S-LFR scheme that achieves the lower convex envelope of the memory-rate pairs in $\{(M_{S},R_{S}): t \in [0:C-r]\} \cup \{(N,0)\}$ where 
		\begin{equation*}
			{(M_{S},R_{S})=\left(\frac{Nt}{C}+\frac{ \binom{C-1}{t+r-1}}{\binom{C}{t}},\frac{\binom{C}{t+r}}{\binom{C}{t}}\right)},
		\end{equation*}
		with subpacketization $\binom{C}{t}$. 
	\end{thm}
	{\it Proof:}
		In the proposed S-LFR scheme above, each file is divided in to $\binom{C}{t}$ sub files. So,  the size of each sub file is $\frac{1}{\binom{C}{t}}$. Each transmission is also having a size of $\frac{1}{\binom{C}{t}}$. For every $t+r$ sized subset of $C$, there is a transmission. So, the total rate $R=\frac{\binom{C}{t+r}}{\binom{C}{t}}$.
		\begin{thm}
			For a shared link Multi-access network with $C$ caches, access degree r, $K$ users, $N$ files there exists an Improved S-LFR scheme that achieves the lower convex envelope of the memory-rate pairs in $\{(M_{IS},R_{IS}): t \in [0:C-r]\} \cup \{(N,0)\}$ where 
			\begin{equation*}
			{(M_{IS},R_{IS})=\left(\frac{Nt}{C}+{ \binom{C-1}{t+r-1}} \left\lceil\frac{F}{r\binom{C}{t}} \right\rceil \frac{1}{F},\frac{\binom{C}{t+r}}{\binom{C}{t}}\right)},
			\end{equation*}
			with subpacketization $\binom{C}{t}$. 
		\end{thm}
	{\it Proof:}
	In the proposed improved S-LFR scheme above, each file is divided in to $\binom{C}{t}$ sub files. So,  the size of each sub file is $\frac{1}{\binom{C}{t}}$. Each transmission is also having a size of $\frac{1}{\binom{C}{t}}$. For every $t+r$ sized subset of $C$, there is a transmission. So, the total rate $R=\frac{\binom{C}{t+r}}{\binom{C}{t}}$.
	\begin{thm}
		For a shared link multi-access cache-aided scalar linear function retrieval problem with $C$ caches, access degree $r$, $K$ users, $N$ files there exists a LFR scheme that achieves the lower convex envelope of the memory-rate pairs in $\{(M_{LFR},R_{LFR}): t \in [0:C-r]\} \cup \{(N,0)\}$ where 
		\begin{equation*}
			(M_{LFR},R_{LFR}) =\left(\frac{tN}{C}, \frac{\binom{C}{t+r}}{\binom{C}{t}} \right),
		\end{equation*}
		with subpacketization $\binom{C}{t}$. 
	\end{thm}
	{\it Proof:}
		If both the security and privacy keys are removed , the SP-LFR scheme degrades to an LFR scheme.
	\begin{thm}
		For a shared link multi-access network with $C$ caches, access degree $r$, $K$ users, $N$ files , the achieved communication load of SP-LFR scheme $R_{SP}(M)$, Improved S-LFR scheme $R_{IS}(M)$ and the worst case optimal rate under uncoded placement $R^{*}(M)$ satisfies \\
		1) At $M=\frac{r\binom{C}{r}}{C}$, $N\geq2Kr$
		\begin{equation*}
		\frac{R_{SP}(M)}{R^{*}(M)} \leq 2
		\end{equation*}
		2) At $M=\frac{\binom{C}{r}}{C}$, $N\geq2K$
			 \begin{equation*}
			 \frac{R_{IS}(M)}{R^{*}(M)} \leq 2
			 \end{equation*}
	\end{thm}
\textit{Proof} : Refer the Appendix
	\begin{exmp}
		Consider the setup considered above with $C$=5, $r$=3, $t$=2, $N$=10, $F = 10 ^{6}$. Number of users , $K=\binom{C}{r}=\binom{5}{3}=10$ and each file is divided in to $\binom{C}{t}=\binom{5}{2}=10$ sub files.  10 users are represented as $\{U_{\{1,2,3\}} ,U_{\{1,2,4\}},U_{\{1,2,5\}} ,U_{\{1,3,4\}},U_{\{1,3,5\}},U_{\{1,4,5\}},U_{\{2,3,4\}}, U_{\{2,3,5\}}, U_{\{2,4,5\}}, U_{\{3,4,5\}}\}$. The sub files are $W_{i,\{1,2\}}, W_{i,\{1,3\}}, W_{i,\{1,4\}}, W_{i,\{1,5\}},
		W_{i,\{2,3\}}, W_{i,\{2,4\}}, W_{i,\{2,5\}}, W_{i,\{3,4\}},
		W_{i,\{3,5\}},\\  W_{i,\{4,5\}}$,$\forall $i$ \in [10]$. \\
		The server generates $ \binom{5}{5} = 1$ security key ${V_{\{1,2,3,4,5\}}}$ independently and uniformly from $\mathcal{F}_{2}^{F/10}$. Then the server generates $K=10$ random vectors as follows :
		\begin{center}
			$\textbf{P}_{g} \triangleq (p_{g,1},...p_{g,10})^{T} \sim Unif\{\mathcal{F}_{2}^{10}\}, \forall g\in \Omega_{2}$.
		\end{center}
		The  privacy keys, denoted by $\{T_{g,\mathcal{T}}; g \in \Omega_{2} , \mathcal{T} \in \Omega_{1},g \cap \mathcal{T}= \emptyset\}$ are generated as in (\ref{eq13}).
		Let
		\begin{center}
			$D_{g,\mathcal{T}} \triangleq T_{g,\mathcal{T}} \oplus V_{g \cup \mathcal{T}} ,\quad  g \in \Omega_{3} , \mathcal{T} \in \Omega_{2},g \cap \mathcal{T}= \emptyset $.
		\end{center}
		Now, Server generates the shares for $D_{g,\mathcal{T}}$ . The 3 shares are denoted as  ${D^{j}_{g,\mathcal{T}} \in \mathcal{F}^{5* 10^{4}}_{4} , j \in [3]}$. $D_{g,\mathcal{T}}$  could be re-constructed from $3$ shares, and we cannot get any information about $D_{g,\mathcal{T}}$ by any $2$ shares or less.
		The contents of each cache is as given below:
		\begin{itemize}
			\item ${Z_{1}=\{W_{i,\{1,2\}},W_{i,\{1,3\}},W_{i,\{1,4\}},W_{i,\{1,5\}} \forall i \in [10]\}} \cup \\ {\{D^{1}_{\{1,2,3\},\{4,5\}},D^{1}_{\{1,2,4\},\{3,5\}},D^{1}_{\{1,2,5\},\{3,4\}},D^{1}_{\{1,3,4\},\{2,5\}},D^{1}_{\{1,3,5\},\{2,4\}}, D^{1}_{\{1,4,5\},\{2,3\}}\}}$.
			\item ${Z_{2}=\{W_{i,\{1,2\}},W_{i,\{2,3\}},W_{i,\{2,4\}},W_{i,\{2,5\}} \forall i \in [10]\}} \cup \\ {\{D^{2}_{\{1,2,3\},\{4,5\}},D^{2}_{\{1,2,4\},\{3,5\}},D^{2}_{\{1,2,5\},\{3,4\}},D^{1}_{\{2,3,4\},\{1,5\}},D^{1}_{\{2,3,5\},\{1,4\}}, D^{1}_{\{2,4,5\},\{1,3\}}\}}$.
			\item ${Z_{3}=\{W_{i,\{1,3\}},W_{i,\{2,3\}},W_{i,\{3,4\}},W_{i,\{3,5\}} \forall i \in [10]\}} \cup \\ {\{D^{3}_{\{1,2,3\},\{4,5\}},D^{2}_{\{1,3,4\},\{2,5\}},D^{2}_{\{1,3,5\},\{2,4\}},D^{2}_{\{2,3,4\},\{1,5\}},D^{2}_{\{2,3,5\},\{1,4\}}, D^{1}_{\{3,4,5\},\{1,2\}}\}}$.
			\item ${Z_{4}=\{W_{i,\{1,4\}},W_{i,\{2,4\}},W_{i,\{3,4\}},W_{i,\{3,5\}} \forall i \in [10]\}} \cup \\ {\{D^{3}_{\{1,2,4\},\{3,5\}},D^{3}_{\{1,3,4\},\{2,5\}},D^{2}_{\{1,4,5\},\{2,3\}},D^{3}_{\{2,3,4\},\{1,5\}},D^{2}_{\{2,4,5\},\{1,3\}}, D^{2}_{\{3,4,5\},\{1,2\}}\}}$.
			\item  ${Z_{5}=\{W_{i,\{1,5\}},W_{i,\{2,5\}},W_{i,\{3,5\}},W_{i,\{4,5\}} \forall i \in [10]\}} \cup \\ {\{D^{3}_{\{1,2,5\},\{3,4\}},D^{3}_{\{1,3,5\},\{2,4\}},D^{3}_{\{1,4,5\},\{2,3\}},D^{3}_{\{2,3,5\},\{1,4\}},D^{3}_{\{2,4,5\},\{1,3\}}, D^{3}_{\{3,4,5\},\{1,2\}}\}}$.
			
		\end{itemize}
		It is clear that the size of each cache is $\frac{46}{10}$.
		Let the users be requesting the following linear combinations:
		$B_{{{\{1,2,3\}}}}= W_{1}+W_{2}$;
		$B_{{{\{1,2,4\}}}}= W_{2}+W_{3}$;
		$B_{{{\{1,2,5\}}}}= W_{3}+W_{4}$;
		$B_{{{\{1,3,4\}}}}= W_{4}+W_{5}$;
		$B_{{{\{1,3,5\}}}}= W_{5}+W_{6}$;
		$B_{{{\{1,4,5\}}}}= W_{6}+W_{7}$;
		$B_{{{\{2,3,4\}}}}= W_{7}+W_{8}$;
		$B_{{{\{2,3,5\}}}}= W_{8}+W_{9}$;
		$B_{{{\{2,4,5\}}}}= W_{9}+W_{10}$;
		$B_{{{\{3,4,5\}}}}= W_{10}$;
		
		Then, the server transmits ${X=[Y_{\{1,2,3,4,5\}},\textbf{q}_{g}]  \quad \forall g \in \Omega^{5}_{3}}$ where
		\begin{itemize}
			\item$Y_{\{1,2,3,4,5\}} =V_{\{1,2,3,4,5\}} \oplus B_{{{\{1,2,3\}}},\{4,5\}} \oplus T_{\{1,2,3\},\{4,5\}} \oplus B_{{{\{1,2,4\}}},\{3,5\}} \oplus T_{\{1,2,4\},\{3,5\}} \oplus B_{{{\{1,2,5\}}},\{3,4\}} \oplus T_{\{1,2,5\},\{3,4\}} \oplus B_{{{\{1,3,4\}}},\{2,5\}} \oplus T_{\{1,3,4\},\{2,5\}} \oplus B_{{{\{1,3,5\}}},\{2,4\}} \oplus T_{\{1,3,5\},\{2,4\}}
			\oplus B_{{{\{1,4,5\}}},\{2,3\}} \oplus T_{\{1,4,5\},\{2,3\}} \oplus B_{{{\{2,3,4\}}},\{1,5\}} \oplus T_{\{2,3,4\},\{1,5\}} \oplus B_{{{\{2,3,5\}}},\{1,4\}} \oplus T_{\{2,3,5\},\{1,4\}} \oplus B_{{{\{2,4,5\}}},\{1,3\}} \oplus T_{\{2,4,5\},\{1,3\}}
			\oplus B_{{{\{3,4,5\}}},\{1,2\}} \oplus T_{\{3,4,5\},\{1,2\}}$,
			\vspace{0.3cm}
			\item $\textbf{q}_{g} = \textbf{P}_{g} + \textbf{d}_{{g}}  \forall g \in \Omega_{3} $. \\
			
		\end{itemize}
		
		Consider the user $U_{\{1,2,3\}}$. The term $B_{{{\{1,2,4\}}},\{3,5\}} \oplus T_{\{1,2,4\},\{3,5\}}$ can be written as  $\underset{i \in [10]}{\bigoplus} {q}_{{\{1,2,4\},i}} W_{i,\{3,5\}}$. As, the user knows $\textbf{q}_{\{1,2,4\}}=({q}_{{\{1,2,4\},1}},.....{q}_{{\{1,2,4\},10}})$ and has access to 3rd cache, it can calculate the term. Similarly, the user $U_{\{1,2,3\}}$ can calculate the terms $B_{{{g}},S\backslash g} \oplus T_{g,S\backslash g} \forall g \subset \{1,2,3,4,5\} , |g|=3, g \neq \{1,2,3\} $. Moreover, the user $U_{\{1,2,3\}}$ can get the key $V_{\{1,2,3,4,5\}} \oplus T_{\{1,2,3\},\{4,5\}} $ from it's caches. so, the user $U_{\{1,2,3\}}$ can decode $B_{{{\{1,2,3\}}},\{4,5\}}$. Similarly all the others can decode the files demanded by them. The privacy is guaranteed since each user does not know the key of the other user, and the vectors $\textbf{q}_{g},  \forall g \in \Omega_{3} $ are random vectors independently and uniformly distributed over $ \mathcal{F}^{10}_{2}$. The security is guaranteed since each transmitted signal is accompanied by a security key. As there is only one transmission, the rate achieved is $\frac{1}{10}$. If we want only security , it can be achieved by using S-LFR scheme given above. The content of each cache is:
		
		\begin{itemize}
			\item ${Z_{1}=\{W_{i,\{1,2\}},W_{i,\{1,3\}},W_{i,\{1,4\}},W_{i,\{1,5\}} \forall i \in [10]\}} \cup  \{V_{\{1,2,3,4,5\}}\}$.
			
			\item ${Z_{2}=\{W_{i,\{1,2\}},W_{i,\{2,3\}},W_{i,\{2,4\}},W_{i,\{2,5\}} \forall i \in [10]\}} \cup  \{V_{\{1,2,3,4,5\}}\}$.
			
			\item ${Z_{3}=\{W_{i,\{1,3\}},W_{i,\{2,3\}},W_{i,\{3,4\}},W_{i,\{3,5\}} \forall i \in [10]\}} \cup  \{V_{\{1,2,3,4,5\}}\}$.
			\item ${Z_{4}=\{W_{i,\{1,4\}},W_{i,\{2,4\}},W_{i,\{3,4\}},W_{i,\{4,5\}} \forall i \in [10]\}} \cup  \{V_{\{1,2,3,4,5\}}\}$.
			
			\item  ${Z_{5}=\{W_{i,\{1,5\}},W_{i,\{2,5\}},W_{i,\{3,5\}},W_{i,\{4,5\}} \forall i \in [10]\}} \cup  \{V_{\{1,2,3,4,5\}}\}$.
			
		\end{itemize}
		From the above placement, the size of each cache is $\frac{41}{10}$.
		In order to satisfy the demands, the server transmits:
		
		$Y_{\{1,2,3,4,5\}} =V_{\{1,2,3,4,5\}} \oplus B_{{{\{1,2,3\}}},\{4,5\}}   \oplus B_{{{\{1,2,4\}}},\{3,5\}}  \oplus B_{{{\{1,2,5\}}},\{3,4\}}  \oplus B_{{{\{1,3,4\}}},\{2,5\}}  \oplus B_{{{\{1,3,5\}}},\{2,4\}} 
		\oplus B_{{{\{1,4,5\}}},\{2,3\}}  \oplus B_{{{\{2,3,4\}}},\{1,5\}}  \oplus B_{{{\{2,3,5\}}},\{1,4\}}   \oplus B_{{{\{2,4,5\}}},\{1,3\}} \oplus 
		\oplus B_{{{\{3,4,5\}}},\{1,2\}} $.\\
		Consider an user $U_{\{1,2,3\}}$. From the placement it is clear that the user $U_{\{1,2,3\}}$  can get all the terms in  $Y_{\{1,2,3,4,5\}}$ except $B_{{{\{1,2,3\}}},\{4,5\}}$ from its cache. So, it gets it's demanded file. Similarly all the other users can get their demanded files. As the transmitted signal is protected by a security key, security is achieved. The rate achieved is $\frac{1}{10}$.
		Now, let us consider the improved SLFR scheme. Do the following for the security key : 
		\begin{itemize}
			\item Divide the key into $3$ sub-keys
			\item Encode them using a (5,3) MDS code.
			\item Now, the coded sub-keys of a key $V_{\{1,2,3,4,5\}}$ are $\tilde{V}_{\{1,2,3,4,5\},1}, \tilde{V}_{\{1,2,3,4,5\},2},\tilde{V}_{\{1,2,3,4,5\},3},\tilde{V}_{\{1,2,3,4,5\},4},\\ \tilde{V}_{\{1,2,3,4,5\},5}$. Any $3$ coded sub-keys of a security key are sufficient to reconstruct the entire key.
		\end{itemize}
	The content of each cache is as follows:
	\begin{itemize}
		\item ${Z_{1}=\{W_{i,\{1,2\}},W_{i,\{1,3\}},W_{i,\{1,4\}},W_{i,\{1,5\}} \forall i \in [10]\}} \cup  \{\tilde{V}_{\{1,2,3,4,5\},1}\}$.
		
		\item ${Z_{2}=\{W_{i,\{1,2\}},W_{i,\{2,3\}},W_{i,\{2,4\}},W_{i,\{2,5\}} \forall i \in [10]\}} \cup  \{\tilde{V}_{\{1,2,3,4,5\},2}\}$.
		
		\item ${Z_{3}=\{W_{i,\{1,3\}},W_{i,\{2,3\}},W_{i,\{3,4\}},W_{i,\{3,5\}} \forall i \in [10]\}} \cup  \{\tilde{V}_{\{1,2,3,4,5\},3}\}$.
		\item ${Z_{4}=\{W_{i,\{1,4\}},W_{i,\{2,4\}},W_{i,\{3,4\}},W_{i,\{4,5\}} \forall i \in [10]\}} \cup  \{\tilde{V}_{\{1,2,3,4,5\},4}\}$.
		
		\item  ${Z_{5}=\{W_{i,\{1,5\}},W_{i,\{2,5\}},W_{i,\{3,5\}},W_{i,\{4,5\}} \forall i \in [10]\}} \cup  \{\tilde{V}_{\{1,2,3,4,5\},5}\}$.
		
	\end{itemize}
The size of coded sub-key is same as the size of sub-key which is $\frac{1}{10}^{th}$ of a file. The size of each cache is $\frac{121}{30}$. In order to satisfy the demands, the server transmits:

$Y_{\{1,2,3,4,5\}} =V_{\{1,2,3,4,5\}} \oplus B_{{{\{1,2,3\}}},\{4,5\}}   \oplus B_{{{\{1,2,4\}}},\{3,5\}}  \oplus B_{{{\{1,2,5\}}},\{3,4\}}  \oplus B_{{{\{1,3,4\}}},\{2,5\}}  \oplus B_{{{\{1,3,5\}}},\{2,4\}} 
\oplus B_{{{\{1,4,5\}}},\{2,3\}}  \oplus B_{{{\{2,3,4\}}},\{1,5\}}  \oplus B_{{{\{2,3,5\}}},\{1,4\}}   \oplus B_{{{\{2,4,5\}}},\{1,3\}} \oplus 
\oplus B_{{{\{3,4,5\}}},\{1,2\}} $.\\
For decodability, consider an user $U_{\{1,2,3\}}$. It has access to $\tilde{V}_{\{1,2,3,4,5\},1}, \tilde{V}_{\{1,2,3,4,5\},2}, \tilde{V}_{\{1,2,3,4,5\},3}$ from its caches. So, it can reconstruct $V_{\{1,2,3,4,5\}}$. It can get  $B_{{{\{1,2,4\}}},\{3,5\}}, B_{{{\{1,3,5\}}},\{2,4\}}, B_{{{\{1,4,5\}}},\{2,3\}},\\ B_{{{\{2,3,4\}}},\{1,5\}}, B_{{{\{2,3,5\}}},\{1,4\}}, B_{{{\{2,4,5\}}},\{1,3\}}, B_{{{\{3,4,5\}}},\{1,2\}}$ as well. So, the user $U_{\{1,2,3\}}$ can get $B_{{{\{1,2,3\}}},\{4,5\}}$. Similarly, all the other users can their demanded files. The rate achived in this scheme is $\frac{1}{10}$. As compared to S-LFR scheme , the improved S-LFR scheme has better performance.

	\end{exmp}

\begin{exmp}
	Consider $C$=5, $r$=2, $t$=2, $N$=10, $F=10 ^{6}$. Number of users , $K=\binom{C}{r}=\binom{5}{2}=10$ and each file is divided in to $\binom{C}{t}=\binom{5}{2}=10$ sub files.  10 users are represented as $\{U_{\{1,2\}} ,U_{\{1,3\}},U_{\{1,4\}} ,U_{\{1,5\}},U_{\{2,3\}},U_{\{2,4\}},U_{\{2,5\}}, U_{\{3,4\}}, U_{\{3,5\}}, U_{\{4,5\}} \}$.. The sub files are $W_{i,\{1,2\}}, W_{i,\{1,3\}}, W_{i,\{1,4\}}, W_{i,\{1,5\}},
	W_{i,\{2,3\}}, W_{i,\{2,4\}}, W_{i,\{2,5\}}, W_{i,\{3,4\}},
	W_{i,\{3,5\}}, W_{i,\{4,5\}}$,$\forall $i$ \in [10]$. \\
	The server generates $ \binom{5}{4} = 5$ security keys \{${V_{S}} : S \in \Omega_{4}$\} independently and uniformly from $\mathcal{F}_{2}^{F/10}$. Then the server generates $K=10$ random vectors as follows :
	\begin{center}
		$\textbf{P}_{g} \triangleq (p_{g,1},...p_{g,10})^{T} \sim Unif\{\mathcal{F}_{2}^{10}\}, \forall g\in \Omega_{2}$.
	\end{center}
	The  privacy keys, denoted by $\{T_{g,\mathcal{T}}; g \in \Omega_{2} , \mathcal{T} \in \Omega_{2},g \cap \mathcal{T}= \emptyset\}$ are generated as in (\ref{eq13}).
	Let
	\begin{center}
		$D_{g,\mathcal{T}} \triangleq T_{g,\mathcal{T}} \oplus V_{g \cup \mathcal{T}} ,\quad  g \in \Omega_{2} , \mathcal{T} \in \Omega_{2},g \cap \mathcal{T}= \emptyset $.
	\end{center}
Now, server generates 2 shares $D^{j}_{g,\mathcal{T}} \in \mathcal{F}^{5* 10^{4}}_{4} , j\in [2]$ for each $D_{g,\mathcal{T}}$ such that $D_{g,\mathcal{T}}$ can be reconstructed using 2 shares and cannot be reconstructed using less than 2 shares. The content of each cache is as follows:
\begin{itemize}
	\item ${Z_{1}=\{W_{i,\{1,2\}},W_{i,\{1,3\}},W_{i,\{1,4\}},W_{i,\{1,5\}} \forall i \in [10]\}} \cup {\{D^{1}_{\{1,2\},\{3,4\}},D^{1}_{\{1,2\},\{3,5\}},D^{1}_{\{1,2\},\{4,5\}}, 
	D^{1}_{\{1,3\},\{2,4\}}}, \\ {D^{1}_{\{1,3\},\{2,5\}}, D^{1}_{\{1,3\},\{4,5\}}, D^{1}_{\{1,4\},\{2,3\}}, D^{1}_{\{1,4\},\{2,5\}}, D^{1}_{\{1,4\},\{3,5\}}, D^{1}_{\{1,5\},\{2,3\}}, D^{1}_{\{1,5\},\{2,4\}}, D^{1}_{\{1,5\},\{3,4\}}\}}$.
	
	\item ${Z_{2}=\{W_{i,\{1,2\}},W_{i,\{2,3\}},W_{i,\{2,4\}},W_{i,\{2,5\}} \forall i \in [10]\}} \cup 
	{\{D^{2}_{\{1,2\},\{3,4\}},D^{2}_{\{1,2\},\{3,5\}},D^{2}_{\{1,2\},\{4,5\}},D^{1}_{\{2,3\},\{1,4\}}}, \\ {D^{1}_{\{2,3\},\{1,5\}}, D^{1}_{\{2,3\},\{4,5\}}, D^{1}_{\{2,4\},\{1,3\}}, D^{1}_{\{2,4\},\{1,5\}}, D^{1}_{\{2,4\},\{3,5\}}, D^{1}_{\{2,5\},\{1,3\}}, D^{1}_{\{2,5\},\{1,4\}}, D^{1}_{\{2,5\},\{3,4\}}\}}$.
		
	\item ${Z_{3}=\{W_{i,\{1,3\}},W_{i,\{2,3\}},W_{i,\{3,4\}},W_{i,\{3,5\}} \forall i \in [10]\}} \cup 
	{\{D^{2}_{\{1,3\},\{2,4\}},D^{2}_{\{1,3\},\{2,5\}},D^{2}_{\{1,3\},\{4,5\}},D^{2}_{\{2,3\},\{1,4\}} },\\ {D^{2}_{\{2,3\},\{1,5\}},  D^{2}_{\{2,3\},\{4,5\}}, D^{1}_{\{3,4\},\{1,3\}}, D^{1}_{\{3,4\},\{1,5\}}, D^{1}_{\{3,4\},\{2,5\}}, D^{1}_{\{3,5\},\{1,2\}}, D^{1}_{\{3,5\},\{1,4\}}, D^{1}_{\{3,5\},\{2,4\}}\}}$.
	
	\item ${Z_{4}=\{W_{i,\{1,4\}},W_{i,\{2,4\}},W_{i,\{3,4\}},W_{i,\{4,5\}} \forall i \in [10]\}} \cup  
	{\{D^{2}_{\{1,4\},\{2,3\}},D^{2}_{\{1,4\},\{2,5\}},D^{2}_{\{1,4\},\{3,5\}},D^{2}_{\{2,4\},\{1,3\}}}, \\ {D^{2}_{\{2,4\},\{1,5\}}, D^{2}_{\{2,4\},\{3,5\}}, D^{2}_{\{3,4\},\{1,3\}}, D^{2}_{\{3,4\},\{1,5\}}, D^{2}_{\{3,4\},\{2,5\}}, D^{1}_{\{4,5\},\{1,2\}}, D^{1}_{\{4,5\},\{1,3\}}, D^{1}_{\{4,5\},\{2,3\}}\}}.$
	
	\item  ${Z_{5}=\{W_{i,\{1,5\}},W_{i,\{2,5\}},W_{i,\{3,5\}},W_{i,\{4,5\}} \forall i \in [10]\}} \cup 
	{\{D^{2}_{\{1,5\},\{2,3\}},D^{2}_{\{1,5\},\{2,4\}},D^{2}_{\{1,5\},\{3,4\}},D^{2}_{\{2,5\},\{1,3\}}}, \\ {D^{2}_{\{2,5\},\{1,4\}}, D^{2}_{\{2,5\},\{3,4\}}, D^{2}_{\{3,5\},\{1,2\}}, D^{2}_{\{3,5\},\{1,4\}}, D^{2}_{\{3,5\},\{2,3\}}, D^{2}_{\{4,5\},\{1,2\}}, D^{2}_{\{4,5\},\{1,3\}}, D^{2}_{\{4,5\},\{2,3\}}\}}.$
	
\end{itemize}
From the above placement, the size of each cache is $\frac{52}{10}$. Let the demand vectors be : $\textbf{d}_{\{1,2\}} = (1 \, 1 \, 1 \, 0 \, 0 \, 0 \, 0 \, 0 \, 0 \, 0 ); \textbf{d}_{\{1,3\}} = (1 \, 1 \, 0 \, 0 \, 0 \, 0 \, 0 \, 1 \, 0 \, 0 ); \textbf{d}_{\{1,4\}} = (1 \, 0 \, 0 \, 0 \, 0 \, 0 \, 0 \, 0 \, 0 \, 1 ); \textbf{d}_{\{1,5\}} = (1 \, 1 \, 1 \, 0 \, 0 \, 0 \, 0 \, 1 \, 1 \, 0 );
\textbf{d}_{\{2,3\}} = (1 \, 0 \, 0 \, 0 \, 0 \, 0 \, 1 \, 1 \, 0 \, 0 );
\textbf{d}_{\{2,4\}} = (1 \, 0 \, 0 \, 0 \, 1 \, 0 \, 1 \, 0 \, 0 \, 0 );
\textbf{d}_{\{2,5\}} = (1 \, 1 \, 0 \, 0 \, 0 \, 1 \, 0 \, 0 \, 0 \, 1 );
\textbf{d}_{\{3,4\}} = (1 \, 1 \, 1 \, 0 \, 0 \, 1 \, 1 \, 0 \, 1 \, 0 );
\textbf{d}_{\{3,5\}} = (0 \, 0 \, 0 \, 0 \, 0 \, 0 \, 0 \, 1 \, 1 \, 0 );
\textbf{d}_{\{4,5\}} = (1 \, 1 \, 0 \, 0 \, 0 \, 0 \, 0 \, 0 \, 0 \, 0 )$.
Then, the server transmits $X=[Y_{\Omega_{4}}, \textbf{q}_{\Omega_{2}}]$ where 
\begin{itemize}
	\item $ Y_{\{1,2,3,4\}}= V_{\{1,2,3,4\}} \oplus B_{\{1,2\},\{3,4\}} \oplus T_{\{1,2\},\{3,4\}}  \oplus B_{\{1,3\},\{2,4\}} \oplus T_{\{1,3\},\{2,4\}} \oplus B_{\{1,4\},\{2,3\}} \oplus T_{\{1,4\},\{2,3\}} \oplus B_{\{2,3\},\{1,4\}} \oplus T_{\{2,3\},\{1,4\}} \oplus B_{\{2,4\},\{1,3\}} \oplus T_{\{2,4\},\{1,3\}} \oplus B_{\{3,4\},\{1,2\}} \oplus T_{\{3,4\},\{1,2\}}$
	\item  $ Y_{\{1,2,3,5\}}= V_{\{1,2,3,5\}} \oplus B_{\{1,2\},\{3,5\}} \oplus T_{\{1,2\},\{3,5\}}  \oplus B_{\{1,3\},\{2,5\}} \oplus T_{\{1,3\},\{2,5\}} \oplus B_{\{1,5\},\{2,3\}} \oplus T_{\{1,5\},\{2,3\}} \oplus B_{\{2,3\},\{1,5\}} \oplus T_{\{2,3\},\{1,5\}} \oplus B_{\{2,5\},\{1,3\}} \oplus T_{\{2,5\},\{1,3\}} \oplus B_{\{3,5\},\{1,2\}} \oplus T_{\{3,5\},\{1,2\}}$
	\item  $ Y_{\{1,2,3,5\}}= V_{\{1,2,3,5\}} \oplus B_{\{1,2\},\{3,5\}} \oplus T_{\{1,2\},\{3,5\}}  \oplus B_{\{1,3\},\{2,5\}} \oplus T_{\{1,3\},\{2,5\}} \oplus B_{\{1,5\},\{2,3\}} \oplus T_{\{1,5\},\{2,3\}} \oplus B_{\{2,3\},\{1,5\}} \oplus T_{\{2,3\},\{1,5\}} \oplus B_{\{2,5\},\{1,3\}} \oplus T_{\{2,5\},\{1,3\}} \oplus B_{\{3,5\},\{1,2\}} \oplus T_{\{3,5\},\{1,2\}}$
	\item  $ Y_{\{1,2,4,5\}}= V_{\{1,2,4,5\}} \oplus B_{\{1,2\},\{4,5\}} \oplus T_{\{1,2\},\{4,5\}}  \oplus B_{\{1,4\},\{2,5\}} \oplus T_{\{1,4\},\{2,5\}} \oplus B_{\{1,5\},\{2,4\}} \oplus T_{\{1,5\},\{2,4\}} \oplus B_{\{2,4\},\{1,5\}} \oplus T_{\{2,4\},\{1,5\}} \oplus B_{\{2,5\},\{1,4\}} \oplus T_{\{2,5\},\{1,4\}} \oplus B_{\{4,5\},\{1,2\}} \oplus T_{\{4,5\},\{1,2\}}$
	\item  $ Y_{\{1,3,4,5\}}= V_{\{1,3,4,5\}} \oplus B_{\{1,3\},\{4,5\}} \oplus T_{\{1,3\},\{4,5\}}  \oplus B_{\{1,4\},\{3,5\}} \oplus T_{\{1,4\},\{3,5\}} \oplus B_{\{1,5\},\{3,4\}} \oplus T_{\{1,5\},\{3,4\}} \oplus B_{\{3,4\},\{1,5\}} \oplus T_{\{3,4\},\{1,5\}} \oplus B_{\{3,5\},\{1,4\}} \oplus T_{\{3,5\},\{1,4\}} \oplus B_{\{4,5\},\{1,3\}} \oplus T_{\{4,5\},\{1,3\}}$
	\item  $ Y_{\{2,3,4,5\}}= V_{\{2,3,4,5\}} \oplus B_{\{2,3\},\{4,5\}} \oplus T_{\{2,3\},\{4,5\}}  \oplus B_{\{2,4\},\{3,5\}} \oplus T_{\{2,4\},\{3,5\}} \oplus B_{\{2,5\},\{3,4\}} \oplus T_{\{2,5\},\{3,4\}} \oplus B_{\{3,4\},\{2,5\}} \oplus T_{\{3,4\},\{2,5\}} \oplus B_{\{3,5\},\{2,4\}} \oplus T_{\{3,5\},\{2,4\}} \oplus B_{\{4,5\},\{2,3\}} \oplus T_{\{4,5\},\{2,3\}}$
	\item $ \textbf{q}_{g} = \textbf{P}_{g} \oplus \textbf{d}_{g} , \forall g \in \Omega_{2}$
\end{itemize}
Consider the user $U_{\{1,2\}}$ and the transmission $Y_{\{1,2,3,4\}}$. It can calculate the terms $B_{{{g}},S\backslash g} \oplus T_{g,S\backslash g} \forall g \subset \{1,2,3,4,\} , |g|=2, g \neq \{1,2\} $ and it can get the key $V_{\{1,2,3,4\}} \oplus T_{\{1,2\},\{3,4\}}$ from its caches. So, it can get $B_{\{1,2\},\{3,4\}}$. Similarly $U_{\{1,2\}}$ can get all the sub files it wants from other transmissions and all the other users can get the sub files they demand. The privacy is guaranteed since each user does not know the key of the other user, and the vectors $\textbf{q}_{g},  \forall g \in \Omega_{2} $ are random vectors independently and uniformly distributed over $ \mathcal{F}^{10}_{2}$. The security is guaranteed since each transmitted signal is accompanied by a security key. If the requirement is only security, then it can be achieved by S-LFR as follows :
The content of each cache is :

\begin{itemize}
	\item ${Z_{1}=\{W_{i,\{1,2\}},W_{i,\{1,3\}},W_{i,\{1,4\}},W_{i,\{1,5\}} \forall i \in [10]\}} \cup  \{V_{\{1,2,3,4\}},V_{\{1,2,3,5\}},V_{\{1,2,4,5\}},V_{\{1,3,4,5\}}\}$.
	\item ${Z_{2}=\{W_{i,\{1,2\}},W_{i,\{2,3\}},W_{i,\{2,4\}},W_{i,\{2,5\}} \forall i \in [10]\}} \cup  \{V_{\{1,2,3,4\}},,V_{\{1,2,3,5\}},V_{\{1,2,4,5\}},V_{\{2,3,4,5\}}\}$.
	\item ${Z_{3}=\{W_{i,\{1,3\}},W_{i,\{2,3\}},W_{i,\{3,4\}},W_{i,\{3,5\}} \forall i \in [10]\}} \cup  \{V_{\{1,2,3,4\}},,V_{\{1,2,3,5\}},V_{\{1,3,4,5\}},V_{\{2,3,4,5\}}\}$.
	\item ${Z_{4}=\{W_{i,\{1,4\}},W_{i,\{2,4\}},W_{i,\{3,4\}},W_{i,\{4,5\}} \forall i \in [10]\}} \cup  \{V_{\{1,2,3,4\}},,V_{\{1,2,4,5\}},V_{\{1,3,4,5\}},V_{\{2,3,4,5\}}\}$.
	\item  ${Z_{5}=\{W_{i,\{1,5\}},W_{i,\{2,5\}},W_{i,\{3,5\}},W_{i,\{4,5\}} \forall i \in [10]\}} \cup  \{V_{\{1,2,3,5\}},,V_{\{1,2,4,5\}},V_{\{1,3,4,5\}},V_{\{2,3,4,5\}}\}$.
	
\end{itemize}
The cache size is $\frac{13}{3}$. In order to satisfy the demands, the server transmits :
\begin{itemize}
	\item $ Y_{\{1,2,3,4\}}= V_{\{1,2,3,4\}} \oplus B_{\{1,2\},\{3,4\}}  \oplus B_{\{1,3\},\{2,4\}}   \oplus B_{\{1,4\},\{2,3\}} \oplus  B_{\{2,3\},\{1,4\}} \oplus  B_{\{2,4\},\{1,3\}}  \oplus B_{\{3,4\},\{1,2\}}$
	\item  $ Y_{\{1,2,3,5\}}= V_{\{1,2,3,5\}} \oplus B_{\{1,2\},\{3,5\}} \oplus  B_{\{1,3\},\{2,5\}}  \oplus B_{\{1,5\},\{2,3\}} \oplus  B_{\{2,3\},\{1,5\}} \oplus  B_{\{2,5\},\{1,3\}} \oplus B_{\{3,5\},\{1,2\}}$
	\item  $ Y_{\{1,2,3,5\}}= V_{\{1,2,3,5\}} \oplus B_{\{1,2\},\{3,5\}}   \oplus B_{\{1,3\},\{2,5\}}  \oplus B_{\{1,5\},\{2,3\}} \oplus B_{\{2,3\},\{1,5\}} \oplus  B_{\{2,5\},\{1,3\}}  \oplus B_{\{3,5\},\{1,2\}}$
	\item  $ Y_{\{1,2,4,5\}}= V_{\{1,2,4,5\}} \oplus B_{\{1,2\},\{4,5\}}   \oplus B_{\{1,4\},\{2,5\}} \oplus  B_{\{1,5\},\{2,4\}}  \oplus B_{\{2,4\},\{1,5\}} \oplus  B_{\{2,5\},\{1,4\}}  \oplus B_{\{4,5\},\{1,2\}} $
	\item  $ Y_{\{1,3,4,5\}}= V_{\{1,3,4,5\}} \oplus B_{\{1,3\},\{4,5\}} \oplus B_{\{1,4\},\{3,5\}}  \oplus B_{\{1,5\},\{3,4\}} \oplus  B_{\{3,4\},\{1,5\}} \oplus  B_{\{3,5\},\{1,4\}} \oplus B_{\{4,5\},\{1,3\}} $
	\item  $ Y_{\{2,3,4,5\}}= V_{\{2,3,4,5\}} \oplus B_{\{2,3\},\{4,5\}}  \oplus B_{\{2,4\},\{3,5\}} \oplus B_{\{2,5\},\{3,4\}} \oplus  B_{\{3,4\},\{2,5\}} \oplus B_{\{3,5\},\{2,4\}} \oplus B_{\{4,5\},\{2,3\}}$
\end{itemize}
Consider the user $U_{\{1,2\}}$ and the transmission $Y_{\{1,2,3,4\}}$. From the placement it is clear that the user $U_{\{1,2\}}$  can get all the terms in  $Y_{\{1,2,3,4\}}$ except $B_{{{\{1,2,\}}},\{3,4\}}$ from its cache. So, using the transmission the user $U_{\{1,2\}}$ gets the sub file $B_{{{\{1,2,\}}},\{3,4\}}$. Similarly it gets the other sub files from other transmissions and all the other users also can get their demanded files. As the transmitted signal is protected by a security key, security is achieved. The rate achieved is $\frac{5}{10}$. Now, lets see the improved S-LFR scheme. Do the following for the security keys : 
\begin{itemize}
	\item Divide each key into $2$ sub-keys
	\item Encode them using a (4,2) MDS code.
	\item Now, there are $4$ coded sub-keys for each key. The coded sub-keys of the keys $\{V_{S}: S \in \Omega_{4}\}$ are as follows :
	\begin{itemize}
		\item The coded sub-keys of $V_{\{1,2,3,4\}}$ are $\tilde{V}_{\{1,2,3,4\},1}, \tilde{V}_{\{1,2,3,4\},2}, \tilde{V}_{\{1,2,3,4\},3},\tilde{V}_{\{1,2,3,4\},4}$.
		\item The coded sub-keys of $V_{\{1,2,3,5\}}$ are $\tilde{V}_{\{1,2,3,5\},1}, \tilde{V}_{\{1,2,3,5\},2}, \tilde{V}_{\{1,2,3,5\},3},\tilde{V}_{\{1,2,3,5\},5}$.
		\item The coded sub-keys of $V_{\{1,2,4,5\}}$ are $\tilde{V}_{\{1,2,4,5\},1}, \tilde{V}_{\{1,2,4,5\},2}, \tilde{V}_{\{1,2,4,5\},4},\tilde{V}_{\{1,2,4,5\},5}$.
		\item The coded sub-keys of $V_{\{1,3,4,5\}}$ are $\tilde{V}_{\{1,3,4,5\},1}, \tilde{V}_{\{1,3,4,5\},3}, \tilde{V}_{\{1,3,4,5\},4},\tilde{V}_{\{1,3,4,5\},5}$.
		\item The coded sub-keys of $V_{\{2,3,4,5\}}$ are $\tilde{V}_{\{2,3,4,5\},2}, \tilde{V}_{\{2,3,4,5\},3}, \tilde{V}_{\{2,3,4,5\},4},\tilde{V}_{\{2,3,4,5\},5}$.
	\end{itemize}
	\item Any $2$ coded sub-keys of a security key are sufficient to reconstruct the entire key.
	
\end{itemize}
Now, the table below shows the way coded sub-keys are placed :

\vspace{0.5cm}
\begin{center}
	\begin{tabular}{ |c|c|c|c|c|}
		\hline
		Cache1 & Cache2 & Cache3 & Cache4 & Cache5 \\
		\hline
		
		$\tilde{V}_{\{1,2,3,4\},1}$ & $\tilde{V}_{\{1,2,3,4\},2}$ & $\tilde{V}_{\{1,2,3,4\},3}$ & $\tilde{V}_{\{1,2,3,4\},4}$ &   \\
		$\tilde{V}_{\{1,2,3,5\},1}$ & $\tilde{V}_{\{1,2,3,5\},2}$ & $\tilde{V}_{\{1,2,3,5\},3}$ &   & $\tilde{V}_{\{1,2,3,5\},5}$   \\ 
		
		$\tilde{V}_{\{1,2,4,5\},1}$ & $\tilde{V}_{\{1,2,4,5\},2}$ &  & $\tilde{V}_{\{1,2,4,5\},4}$   & $\tilde{V}_{\{1,2,4,5\},5}$   \\  
		
		$\tilde{V}_{\{1,3,4,5\},1}$ &   & $\tilde{V}_{\{1,3,4,5\},3}$  & $\tilde{V}_{\{1,3,4,5\},4}$   & $\tilde{V}_{\{1,3,4,5\},5}$   \\  
		
		   & $\tilde{V}_{\{2,3,4,5\},2}$  & $\tilde{V}_{\{2,3,4,5\},3}$  & $\tilde{V}_{\{2,3,4,5\},4}$   & $\tilde{V}_{\{2,3,4,5\},5}$   \\
		   \hline
		   
	\end{tabular}
\end{center}
\vspace{0.5cm}
From the above table, it is clear that each user have access to $2$ coded sub-keys of the security key it wanted. For example, the user $U_{\{1,2\}}$ has access to $2$ coded sub-keys of the keys $V_{\{1,2,3,4\}}, V_{\{1,2,3,5\}}, V_{\{1,2,4,5\}}, V_{\{1,3,4,5\}}$. With $2$ coded sub-keys, the user can reconstruct the entire key. The overall cache placement is as follows :
\begin{itemize}
	\item ${Z_{1}=\{W_{i,\{1,2\}},W_{i,\{1,3\}},W_{i,\{1,4\}},W_{i,\{1,5\}} \forall i \in [10]\}} \cup  \{\tilde{V}_{\{1,2,3,4\},1},\tilde{V}_{\{1,2,3,5\},1}, \tilde{V}_{\{1,2,4,5\},1}, \tilde{V}_{\{1,3,4,5\},1}\}$.
	\item ${Z_{2}=\{W_{i,\{1,2\}},W_{i,\{2,3\}},W_{i,\{2,4\}},W_{i,\{2,5\}} \forall i \in [10]\}} \cup  \{\tilde{V}_{\{1,2,3,4\},2},\tilde{V}_{\{1,2,3,5\},2}, \tilde{V}_{\{1,2,4,5\},2}, \tilde{V}_{\{2,3,4,5\},2} \}$.
	\item ${Z_{3}=\{W_{i,\{1,3\}},W_{i,\{2,3\}},W_{i,\{3,4\}},W_{i,\{3,5\}} \forall i \in [10]\}} \cup  \{\tilde{V}_{\{1,2,3,4\},3},\tilde{V}_{\{1,2,3,5\},3}, \tilde{V}_{\{1,3,4,5\},3}, \tilde{V}_{\{2,3,4,5\},3} \}$.
	\item ${Z_{4}=\{W_{i,\{1,4\}},W_{i,\{2,4\}},W_{i,\{3,4\}},W_{i,\{4,5\}} \forall i \in [10]\}} \cup  \{\tilde{V}_{\{1,2,3,4\},4},\tilde{V}_{\{1,2,4,5\},4}, \tilde{V}_{\{1,3,4,5\},4}, \tilde{V}_{\{2,3,4,5\},4} \}$.
	\item  ${Z_{5}=\{W_{i,\{1,5\}},W_{i,\{2,5\}},W_{i,\{3,5\}},W_{i,\{4,5\}} \forall i \in [10]\}} \cup  \{\tilde{V}_{\{1,2,3,5\},5},\tilde{V}_{\{1,2,4,5\},5}, \tilde{V}_{\{1,3,4,5\},5}, \tilde{V}_{\{2,3,4,5\},5} \}$.
	
\end{itemize}
The cache size is $\frac{13}{3}$. In order to satisfy the demands, the server transmits :
\begin{itemize}
	\item $ Y_{\{1,2,3,4\}}= V_{\{1,2,3,4\}} \oplus B_{\{1,2\},\{3,4\}}  \oplus B_{\{1,3\},\{2,4\}}   \oplus B_{\{1,4\},\{2,3\}} \oplus  B_{\{2,3\},\{1,4\}} \oplus  B_{\{2,4\},\{1,3\}}  \oplus B_{\{3,4\},\{1,2\}}$
	\item  $ Y_{\{1,2,3,5\}}= V_{\{1,2,3,5\}} \oplus B_{\{1,2\},\{3,5\}} \oplus  B_{\{1,3\},\{2,5\}}  \oplus B_{\{1,5\},\{2,3\}} \oplus  B_{\{2,3\},\{1,5\}} \oplus  B_{\{2,5\},\{1,3\}} \oplus B_{\{3,5\},\{1,2\}}$
	\item  $ Y_{\{1,2,3,5\}}= V_{\{1,2,3,5\}} \oplus B_{\{1,2\},\{3,5\}}   \oplus B_{\{1,3\},\{2,5\}}  \oplus B_{\{1,5\},\{2,3\}} \oplus B_{\{2,3\},\{1,5\}} \oplus  B_{\{2,5\},\{1,3\}}  \oplus B_{\{3,5\},\{1,2\}}$
	\item  $ Y_{\{1,2,4,5\}}= V_{\{1,2,4,5\}} \oplus B_{\{1,2\},\{4,5\}}   \oplus B_{\{1,4\},\{2,5\}} \oplus  B_{\{1,5\},\{2,4\}}  \oplus B_{\{2,4\},\{1,5\}} \oplus  B_{\{2,5\},\{1,4\}}  \oplus B_{\{4,5\},\{1,2\}} $
	\item  $ Y_{\{1,3,4,5\}}= V_{\{1,3,4,5\}} \oplus B_{\{1,3\},\{4,5\}} \oplus B_{\{1,4\},\{3,5\}}  \oplus B_{\{1,5\},\{3,4\}} \oplus  B_{\{3,4\},\{1,5\}} \oplus  B_{\{3,5\},\{1,4\}} \oplus B_{\{4,5\},\{1,3\}} $
	\item  $ Y_{\{2,3,4,5\}}= V_{\{2,3,4,5\}} \oplus B_{\{2,3\},\{4,5\}}  \oplus B_{\{2,4\},\{3,5\}} \oplus B_{\{2,5\},\{3,4\}} \oplus  B_{\{3,4\},\{2,5\}} \oplus B_{\{3,5\},\{2,4\}} \oplus B_{\{4,5\},\{2,3\}}$
\end{itemize}
Consider the user $U_{\{1,2\}}$ and the transmission $Y_{\{1,2,3,4\}}$. The $U_{\{1,2\}}$ needs to get the sub files $B_{\{1,2\},\{3,4\}}, B_{\{1,2\},\{3,5\}}, B_{\{1,2\},\{4,5\}}$. It can reconstruct $V_{\{1,2,3,4\}}$ as it has access to $2$ coded sub files $\tilde{V}_{\{1,2,3,4\},1}$ and $\tilde{V}_{\{1,2,3,4\},2}$. It can also get the sub files $B_{\{1,3\},\{2,4\}}, B_{\{1,4\},\{2,3\}}, B_{\{2,3\},\{1,4\}}, B_{\{2,4\},\{1,3\}}, \\ B_{\{3,4\},\{1,2\}}$. So, it can get $B_{\{1,2\},\{3,4\}}$ from the transmission $Y_{\{1,2,3,4\}}$. Similarly it gets $B_{\{1,2\},\{3,5\}}$ from $Y_{\{1,2,3,4\}}$, $B_{\{1,2\},\{4,5\}}$ from $Y_{\{1,2,4,5\}}$. Similarly, all the other users also can get their demanded files. As the transmitted signal is protected by a security key, security is achieved. The rate achieved is $\frac{5}{10}$.

\end{exmp}

	\section{CONCLUSION}
	For the multi-access network in \cite{MKR}, we investigated content security and demand privacy when each user is interested in decoding linear combination of the files. We proposed an achievable Secure, Private LFR (SP-LFR) scheme from which Private LFR (P-LFR) scheme, LFR schemes can be derived. We also proposed an achievable Secure LFR (S-LFR) scheme, Improved S-LFR scheme. At a memory point, the SP-LFR and Improved S-LFR scheme are with in a constant multiplicative gap when $N\geq 2Kr$ and $N\geq2K$ respectively. We compared the performances of SP-LFR, P-LFR, S-LFR, Improved S-LFR, LFR schemes with that of SFR scheme in \cite{MKR}.
	\section*{Acknowledgment}
	This work was supported partly by the Science and Engineering Research Board (SERB) of Department of Science and Technology (DST), Government of India, through J.C. Bose National Fellowship to B. Sundar Rajan.
	
	\appendix 
	\begin{center}
		\bf{Proof of Theorem 6}
	\end{center}
	
	Consider a multi-access network consisting of  a server that contain $N$ files which is connected to $K$ users through a error free shared link. Each user is having an access to a unique set of $r$ caches out of $C$ caches. $R^{*}(M)$ represents the worst-case optimal rate with out security under the assumption of uncoded placement of file contents.\\
	1) At $M=\frac{r\binom{C}{r}}{C}$, we need to show the optimality. For the SP-LFR scheme this memory point occurs when $t=0$. Thus,
	\begin{equation}
	R_{SP}\left(M=\frac{r\binom{C}{r}}{C}\right)=\binom{C}{r}
	\end{equation} 
	Let us assume that $N>2rK$ . From theorem 1 in \cite{MKR}, we get $R^{*}(M=\frac{r\binom{C}{r}}{C})$ by doing memory sharing between the points $(0,K)$ and $(\frac{N}{C},\frac{\binom{C}{r+1}}{C})$. We obtain,
	\begin{equation}
	R^{*}\left({M=\frac{r\binom{C}{r}}{C}}\right)= K-\frac{K^{2}r}{N}\left(1-\frac{C-r}{C(r+1}\right)
	\end{equation}
	Therefore, at $M=\frac{r\binom{C}{r}}{C}$
	\begin{subequations}
		\begin{align*}
		\frac{R_{SP}(M)}{R^{*}(M)}&=\frac{K}{K-\frac{K^{2}r}{N}\left(1-\frac{C-r}{C(r+1)}\right)}\\
		&= \frac{K}{K-\frac{rK^{2}}{N}\left(\frac{Cr+r}{Cr+C}\right)}
		\end{align*}
	\end{subequations}
	As $r\leq c$, $K-\frac{rK^{2}}{N} \leq K-\frac{rK^{2}}{N}\left(\frac{Cr+r}{Cr+C}\right)$ . So, we obtain
	\begin{subequations}
		\begin{align*}
		\frac{R_{IS}(M)}{R^{*}{M}} &\leq \frac{K}{K-\frac{K^{2}r}{N}} \\
		&= \frac{1}{1-\frac{Kr}{N}} \\
		&\leq 2 \quad  for N\geq 2Kr .
		\end{align*}
	\end{subequations}
	
	2) At $M=\frac{\binom{C}{r}}{C}$, we need to show the optimality. For the Improved S-LFR scheme this memory point occurs when $t=0$. Thus,
	\begin{equation}
	R_{IS}\left(M=\frac{\binom{C}{r}}{C}\right)=\binom{C}{r}
	\end{equation} 
	From theorem 1 in \cite{MKR}, we get $R^{*}(M=\frac{\binom{C}{r}}{C})$ by doing memory sharing between the points $(0,K)$ and $(\frac{N}{C},\frac{\binom{C}{r+1}}{C})$. We obtain,
	\begin{equation}
	R^{*}\left({M=\frac{\binom{C}{r}}{C}}\right)= K-\frac{K^{2}}{N}\left(1-\frac{C-r}{C(r+1}\right)
	\end{equation}
	Therefore, at $M=\frac{\binom{C}{r}}{C}$
	\begin{subequations}
		\begin{align*}
		\frac{R_{IS}(M)}{R^{*}{M}}&=\frac{K}{K-\frac{K^{2}}{N}\left(1-\frac{C-r}{C(r+1)}\right)}\\
		&= \frac{K}{K-\frac{K^{2}}{N}\left(\frac{Cr+r}{Cr+C}\right)}
		\end{align*}
	\end{subequations}
	As $r\leq c$, $K-\frac{K^{2}}{N} \leq K-\frac{K^{2}}{N}\left(\frac{Cr+r}{Cr+C}\right)$ . So, we obtain
	\begin{subequations}
		\begin{align*}
		\frac{R_{IS}(M)}{R^{*}{M}} &\leq \frac{K}{K-\frac{K^{2}}{N}} \\
		&= \frac{1}{1-\frac{K}{N}} \\
		&\leq 2 \quad  for N\geq 2K .
		\end{align*}
	\end{subequations}

	


\end{document}